\documentclass[twocolumn,epjc3]{svjour3}
\usepackage{graphics}
\usepackage{multicol}
\usepackage{cuted}
\usepackage{flushend}
\usepackage{microtype}
\usepackage{bbm}
\usepackage{amsmath}
\usepackage{mathtools}
\usepackage{caption}
\usepackage{subcaption}
\usepackage{amssymb}
\usepackage{mathrsfs}       
\usepackage[pdftex]{graphicx}
\usepackage{pgfplots}
\pgfplotsset{width=10cm,compat=1.9}
\usepackage{xargs}       
\usepackage{wasysym}
\usepackage{engrec}
\usepackage{enumerate}
\usepackage{appendix}
\usepackage{empheq}
\usepackage{float}
\usepackage{stmaryrd}
\usepackage[parfill]{parskip}
\usepackage[pdftex,breaklinks]{hyperref}
\usepackage{titletoc}
\usepackage{makecell}
\usepackage{lscape}
\usepackage{algorithm}
\usepackage{algorithmicx}
\usepackage{tensor}
\usepackage{algpseudocode}
\usepackage{blkarray}
\usepackage{multirow}
\usepackage{bigstrut}
\usetikzlibrary{plotmarks}
\usepackage{comment}
\usepackage{cite}

\definecolor{mbscolor}{rgb}{0.60, 0.0, 0.65}

\newcommand{\Au}{$^{197}$Au}

\newenvironment{customlegend}[1][]{%
    \begingroup
    \csname pgfplots@init@cleared@structures\endcsname
    \pgfplotsset{#1}%
}{%
    \csname pgfplots@createlegend\endcsname
    \endgroup
}%
\def\addlegendimage{\csname pgfplots@addlegendimage\endcsname}

\newcommand{\trento}{T$\mathrel{\protect\raisebox{-2.1pt}{R}}$ENTo}
 
\begin{document}
\title{The shape of gold}

\author{B.~Bally\thanksref{ad:esnt,ad:dft,em:bb} \and G.~Giacalone\thanksref{ad:itp,em:gg} \and M.~Bender\thanksref{ad:ipnl,em:mb}} 

\thankstext{em:bb}{\email{benjamin.bally@cea.fr}}
\thankstext{em:gg}{\email{giacalone@thphys.uni-heidelberg.de}}
\thankstext{em:mb}{\email{bender@ip2i.in2p3.fr}}

\institute{
\label{ad:esnt}
ESNT, IRFU, CEA, Universit\'e Paris-Saclay, 91191 Gif-sur-Yvette, France 
\and
\label{ad:dft}
Departamento de F\'isica Te\'orica, Universidad Aut\'onoma de Madrid, E-28049 Madrid, Spain
\and
\label{ad:itp}
Institut f\"ur Theoretische Physik, Universit\"at Heidelberg, Philosophenweg 16, D-69120 Heidelberg, Germany
\and
\label{ad:ipnl}
Universit{\'e} de Lyon,
Universit{\'e} Claude Bernard Lyon 1, 
CNRS/IN2P3, 
IP2I Lyon, UMR 5822, 
4 rue Enrico Fermi, 
F-69622, Villeurbanne, France
}

\date{Received: \today{} / Revised version: date}

\maketitle

%
%
\begin{abstract}
Having a detailed theoretical knowledge of the low-energy structure of the heavy odd-mass nucleus $^{197}$Au is of prime interest as the structure of this isotope represents an important input to theoretical simulations of collider experiments involving gold ions performed worldwide at relativistic energies. In the present article, therefore, we report on new results on the structure of $^{197}$Au obtained from state-of-the-art multi-reference energy density functional (MR-EDF) calculations. Our MR-EDF calculations were realized using the Skyrme-type pseudo-potential SLyMR1, and include beyond mean-field correlations through the mixing, in the spirit of the Generator Coordinate Method (GCM), of particle-number and angular-momentum projected triaxially deformed Bogoliubov quasi-particle states. Comparison with experimental data shows that the model gives a reasonable description of \Au~with in particular a good agreement for most of the spectroscopic properties of the $3/2_1^+$ ground state. From the collective wave function of the correlated state, we compute an average deformation $\bar{\beta}(3/2_1^+)=0.13$ and $\bar{\gamma}(3/2_1^+)=40^\circ$ for the ground state. We use this result to construct an intrinsic shape of $^{197}$Au representing a microscopically-motivated input for precision simulations of the associated collider processes. We discuss, in particular, how the triaxiality of this nucleus is expected to impact $^{197}$Au+$^{197}$Au collision experiments at ultrarelativistic energy.
\end{abstract}
%
%
\section{Introduction}
\label{sec:intro}

For millennia, gold has held a prominent role in human societies, whether it be as a symbol of wealth, a standard in international economic trades or because of its medicinal and industrial applications. Interestingly, all the gold of the world, whether it is used as jewelry, in computer chips or kept in secured bank vaults, shares one important feature: it is made of a single isotope. Indeed, zooming in on the structure of this special element at the nuclear scale, one discovers that there is only one stable gold isotope known to exist, namely $^{197}$Au.

As a matter of fact, nuclear physics essentially began with the $^{197}$Au nucleus, which has been the first to be discovered in 1909 by Rutherford, Geiger and Mardsen from the scattering of $\alpha$ particles off a gold foil \cite{Geiger09a,Rutherford11a}. Over 100 years later, we have now a wealth of data available on the structure of $^{197}$Au \cite{Huang05a,Elliott39a,Cook54a,Bolotin79a,Stuchbery88a,Fotiades05a,Wheldon06a,Fotiades09a}. The low-energy spectrum of the nucleus is well known and electromagnetic moments were measured for the ground state as well as for several excited states \cite{Stone19a,Stone20a}. Within a simple single-particle picture, the $3/2_1^+$ ground state of  $^{197}$Au can be interpreted as a proton $2\text{d}_{3/2}$ particle (hole) coupled to a $^{196}$Pt ($^{198}$Hg) core.
Considering the naive picture of a many-body state built as the product of independent harmonic oscillator single-particles (holes) on top of a suitably chosen core, oblate deformations are favoured for nuclei close to the end of a major shell \cite{BM98a,Hamamoto09a}.
Given the proximity of the $Z=82$ and $N=126$ shell closures, we can thus expect \Au~to adopt a small oblate-like deformation. Actually, axially-symmetric mean-field calculations based on the Gogny D1S functional \cite{Decharge80a,Berger91a} reported in the AMEDEE database \cite{Hilaire07a} do find an oblate minimum with a magnitude of $\beta \approx 0.12$.

Starting from the early 2000's, $^{197}$Au has played a central role as well in high-energy nuclear physics. Indeed, gold ions are employed in various scattering experiments ranging from fixed-target experiments at a nucleon-nucleon center-of-mass energy of 2-3 GeV performed at GSI, Darmstadt, to ultrarelativsitic collisions at a nucleon-nucleon center-of-mass energy of 200 GeV performed at the at the BNL Relativsitic Heavy Ion Collider (RHIC). Gold is, in particular, the prime species used at the BNL RHIC, and the first conclusive evidence of the formation of quark-gluon plasma in a laboratory has been indeed obtained in  ultrarelativistic $^{197}$Au+$^{197}$Au collisions \cite{Adams05a,Adcox05a,Back05a,Arsene05a}.

The theoretical interpretation of the results of high-energy scattering experiments starts with an input from nuclear structure theory \cite{Miller07a}. The great success of the hydrodynamic modeling of the quark-gluon plasma
\cite{Busza18a} combined with the availability of data from collisions of several ion species has recently lead to a precise identification of the impact of the structural properties of the collided nuclei on several experimental observables. In particular, the azimuthal distributions of particles produced in relativistic collision experiments are observed to present a strong sensitivity to spatial correlations of nucleons (i.e. deformations) in the ground-state many-body wave function of the colliding species \cite{Adamczyk15a,ALICE:2018lao,Sirunyan19a,Aad20a,Abdallah22b,Townhall22a}. For example, in a recent article \cite{Bally22a}, we argued that we could identify fingerprints of the triaxiality of $^{129}$Xe in collisions performed at the CERN Large Hadron Collider (LHC). The picture of a triaxial $^{129}$Xe drawn from the analysis of high-energy data \cite{ATLAS:2022dov} is in excellent agreement with results obtained from low-energy Coulomb excitation experiments performed on the adjacent isotopes, $^{128,130}$Xe \cite{Morrison20a,Kisyov22a}, as well as with our recent theoretical calculations dedicated to these three xenon isotopes \cite{Bally22b}. Our goal for this manuscript is, in a sense, to perform a similar analysis focused on \Au{}, to assess and potentially improve the current structure input to high-energy $^{197}$Au+$^{197}$Au collisions.

To this aim, we first investigate the low-energy structure of \Au{} on  microscopic grounds using the MR-EDF formalism \cite{Bender03a,Schunck19}. More precisely, we present new results obtained from state-of-the-art calculations based on the configuration mixing of symmetry-projected triaxially deformed Bogoliubov quasi-particle states \cite{Bender08a,Rodriguez10a,Yao10a,Bally14a,Borrajo17a,Bally21a} and the use of the Skyrme-type pseudo-potential SLyMR1 \cite{Sadoudi13b,JodonPHD}. Secondly, we employ these results to construct a point-nucleon density for $^{197}$Au, which we subsequently employ in state-of-the-art simulations of the initial states of high-energy $^{197}$Au+$^{197}$Au collisions. We point out, thus, the expected consequences of implementing our newly-derived nucleon density in future hydrodynamic simulations of such processes, with a focus on the role played by the presence of a slight triaxiality in the colliding ions.

This article is organized as follows: In Sec.~\ref{sec:nucstru}, we report on MR-EDF calculations dedicated to the study of the structure of $^{197}$Au. Then, in Sec.~\ref{sec:collisions} we analyze the consequences of our results on the modeling and the observables of relativistic heavy-ion collisions. Finally, our conclusions and prospects are reported in Sec.~\ref{sec:conclu}.

%
%
\section{Nuclear structure}
\label{sec:nucstru}

\subsection{Method}
\label{sec:method}

In the present study, we use the same theoretical framework as the one that was presented in Ref.~\cite{Bally22b} and refer to that article for more details on our method such as the definitions of the usual operators or the symmetries used in our calculations. Nevertheless, to deal with the heavy-mass \Au~nucleus we changed a few numerical parameters compared to the ones used in Ref.~\cite{Bally22b}. Firstly, the Bogoliubov reference states were represented on a three-dimensional Cartesian Lagrange mesh \cite{Baye86a} in a box of 32 points in each direction. Secondly, when exploring the triaxial deformations, we used a mesh with a spacing\footnote{When considering axial deformations, this corresponds to a step of $\Delta \beta \approx 0.05$.} $\Delta q_1 = \Delta q_2 = 375$ fm$^2$ starting from $(q_1,q_2) = (0,0)$ and restricting ourselves to positive values of $q_1$ and $q_2$, which maps the first sextant of the $\beta$-$\gamma$ plane. Finally, concerning the cutoffs applied during the mixing of reference states: before the mixing, we remove the projected components that in the decomposition of the original reference states have a weight that is lower than $10^{-3}$, whereas during the mixing of $K$-components (performed individually for each reference state) we remove the norm eigenstates with an eigenvalue smaller than $10^{-2}$, and during the final diagonalization mixing projected states originating from different Bogoliubov vacua, we remove the norm eigenstates with an eigenvalue smaller than $10^{-4}$ for all nuclei. The values for the the cutoffs are more restrictive than the ones used when tackling the $^{128,129,130}$Xe isotopes because the configuration mixing performed in the present calculations for $^{197}$Au proved to be more sensitive to the inclusion of components with a small weight that are probably not well represented on our cartesian mesh and, therefore, have to be discarded.
Unfortunately, the improvement of the numerical accuracy of our lattice, by increasing the number of mesh points and/or reducing the spacing between them, implies a substantial increase of the computational cost of the MR-EDF calculations that is at present out of reach for us.

\subsection{Structure of $^{197}$Au}
\label{sec:structure}

\subsubsection{Energy surfaces}
\label{sec:surface}

\begin{figure}[t!]
    \centering
    \includegraphics[width=.70\linewidth]{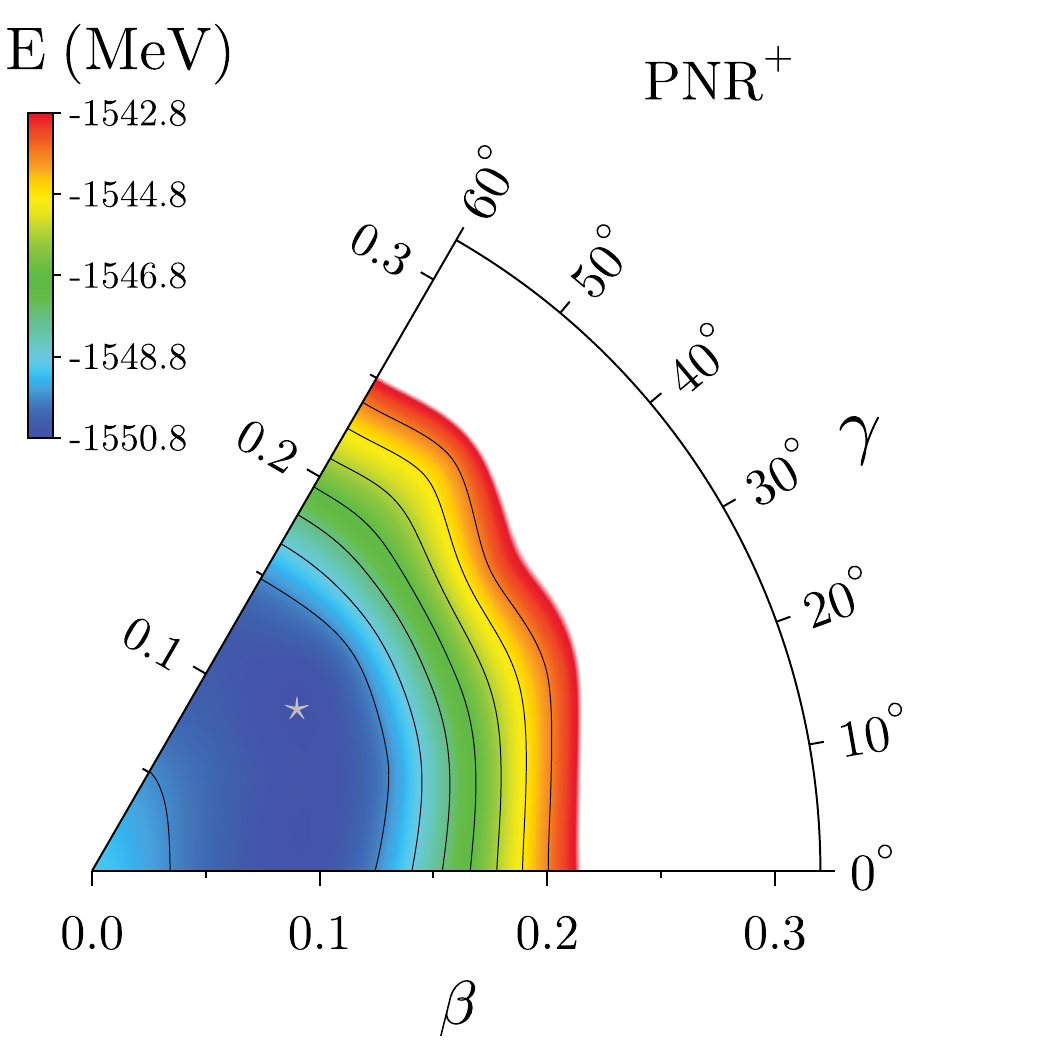}     \includegraphics[width=.70\linewidth]{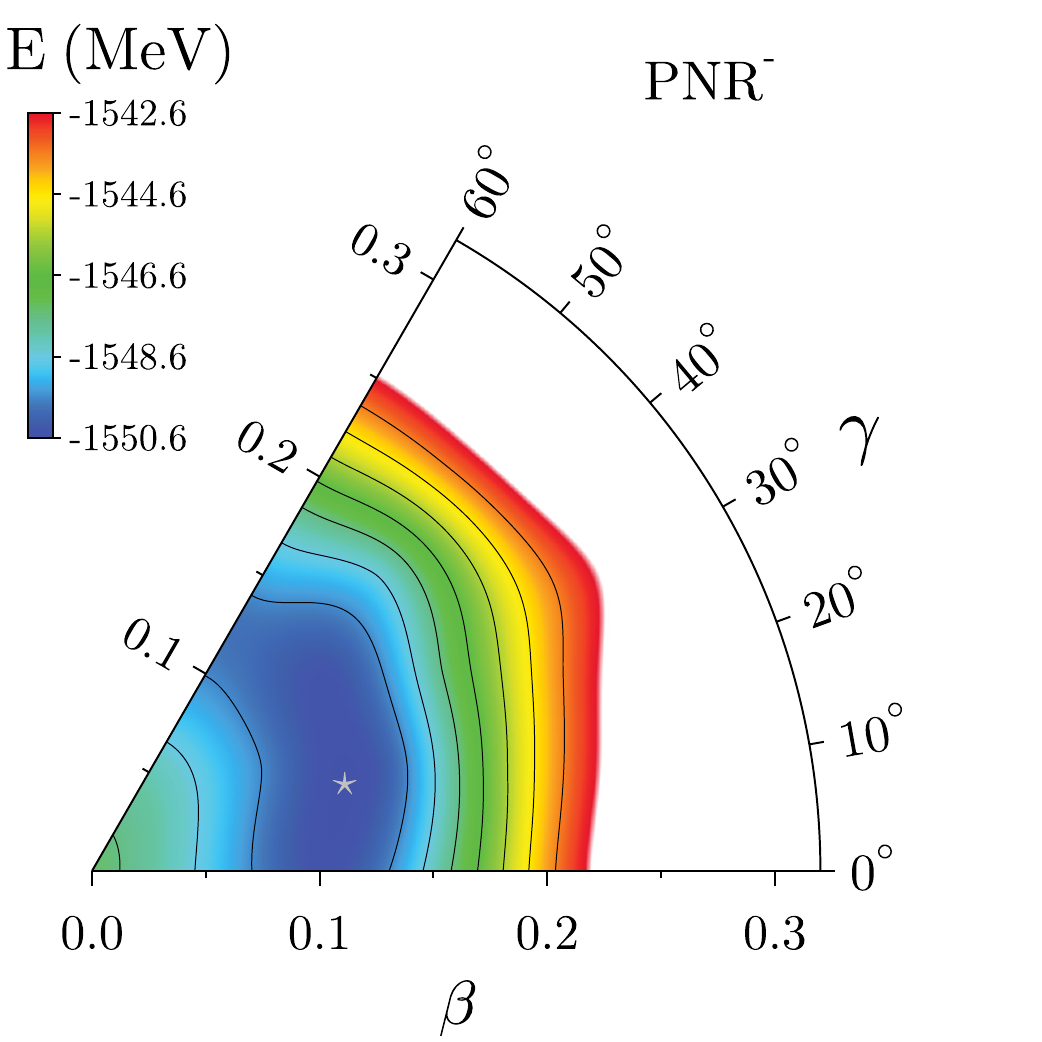}  
    \caption{Particle-number restored total energy surfaces for $^{197}$Au and $\pi=+1$ (top panel) or $\pi=-1$ (bottom panel). Black lines are separated by 1 MeV.
    The minimum for positive (negative) parity, indicated by a silver star, is located at a deformation of $\beta=0.12$ and $\gamma=38^\circ$ ($\beta=0.12$ and $\gamma=19^\circ$)}
     \label{fig:au197_ener_pnr}
\end{figure}

The first step in our approach is the generation of a set of one-quasi-particle states that will be used as reference states in the final configuration mixing calculations. To generate and select the reference states, we follow the strategy detailed in Ref.~\cite{Bally22b}. We briefly recall here that this implies: i) the self-consistent blocking of four different one-quasi-particle states at each point of the deformation mesh, ii) the projection onto good particle numbers and good angular momentum of all (converged) one-quasi-particle states, and iii) the selection of the ones having a projected energy lower than a given threshold above the projected minimum of same parity. In this work, we use a threshold of 5 MeV for both positive and negative parity states.

But before discussing the final results obtained after configuration mixing, let us first analyze the intermediate steps in our method. Figure~\ref{fig:au197_ener_pnr} displays the particle-number restored (PNR) total energy surface for the positive and negative parity states of $^{197}$Au. As can be seen, the two energy surfaces exhibit a $\gamma$-soft topography with a slightly deformed minimum\footnote{Note that all the extrema discussed in this article are computed from an interpolation based on the results obtained at the points on the discretized deformation mesh.} located at $\beta=0.12$. Also, we notice that the surface for positive parity is softer at small deformation than the surface for negative parity. Finally, the minimum for positive parity states is approximately 200 keV lower than the one for negative parity states.

\begin{figure}[t!]
    \centering
    \includegraphics[width=.70\linewidth]{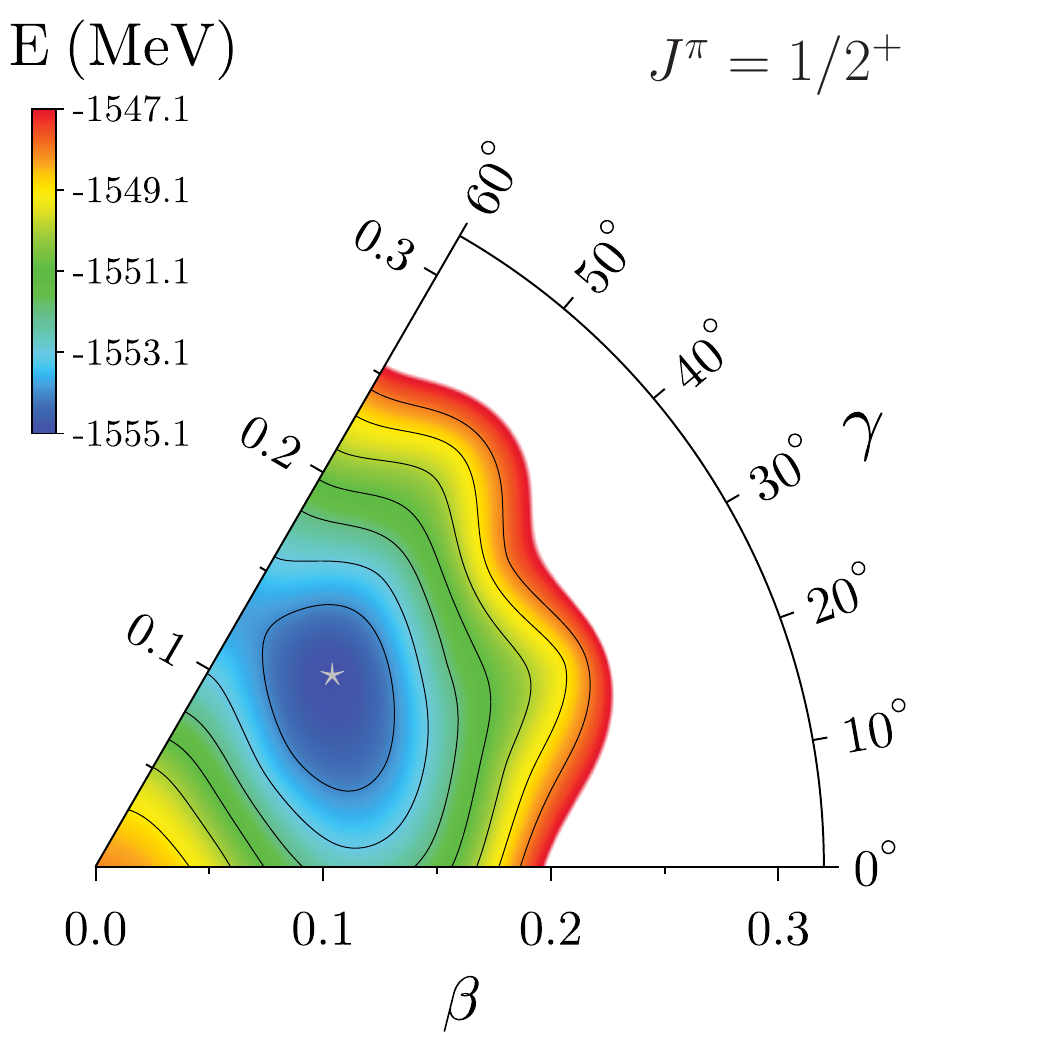} 
    \includegraphics[width=.70\linewidth]{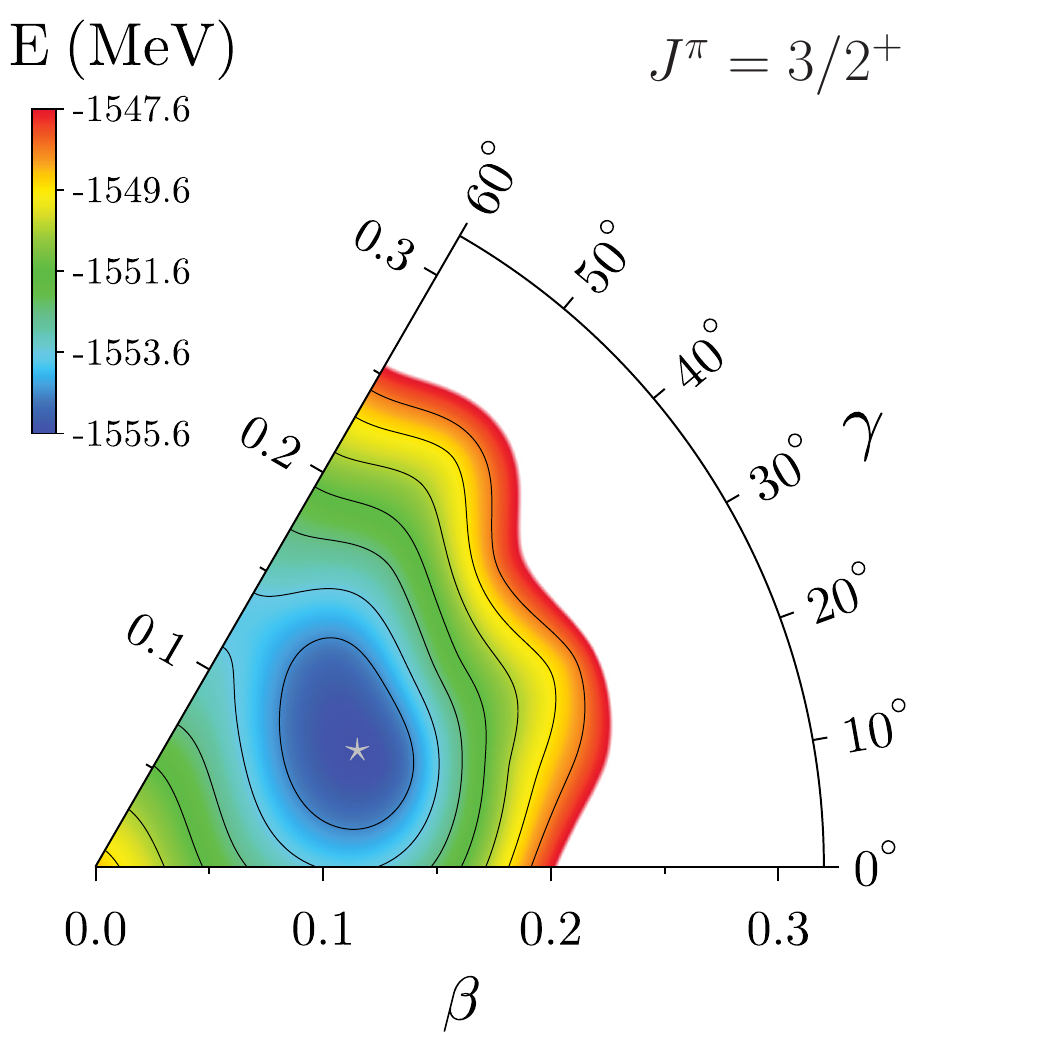}
    \includegraphics[width=.70\linewidth]{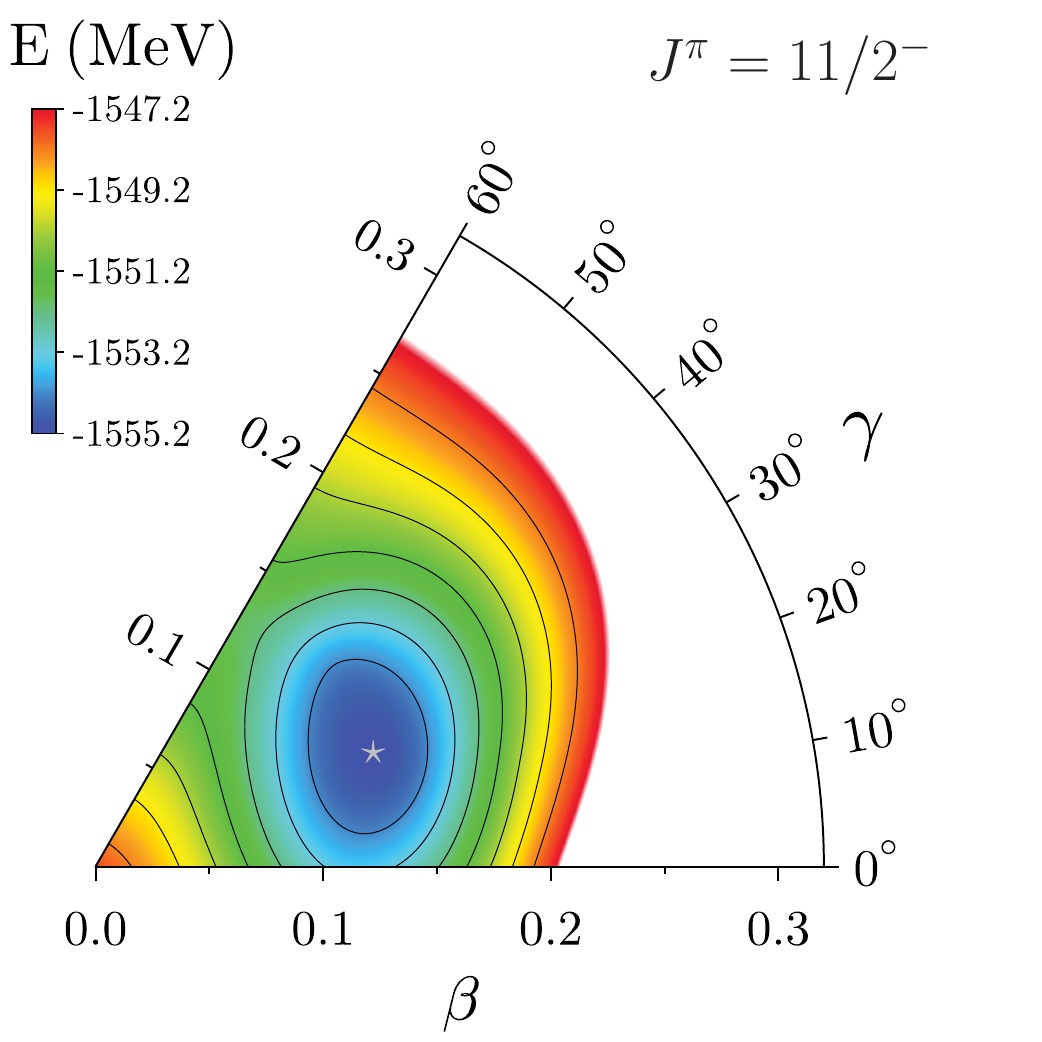} 
    \caption{Angular-momentum and particle-number restored total energy surface for $^{197}$Au and for the lowest $J^\pi = 1/2^+$ (top panel), the lowest $J^\pi = 3/2^+$ (middle panel) and lowest  $J^\pi = 11/2^-$ (bottom panel). Black lines are separated by 1 MeV. The minima for $J^\pi = 1/2^+$, $3/2^+$ and $11/2^-$, indicated by silver stars, are located at deformations of $\beta=0.13$ and $\gamma=39^\circ$, $\beta=0.13$ and $\gamma=24^\circ$ and $\beta=0.13$ and $\gamma=22^\circ$, respectively}
    \label{fig:au197_ener_ampnr_all}
\end{figure}

Performing the full symmetry restoration, we display in Fig.~\ref{fig:au197_ener_ampnr_all}  the angular-momentum and particle-number restored (AMPNR) total energy surfaces for the lowest $J^\pi = 1/2^+$, $3/2^+$ and $11/2^-$ projected states, which are the three values of $J^\pi$ giving the lowest projected energies.
A first remark is that the energy surfaces are much more rigid with now a well pronounced triaxial minimum with $\beta=0.13$. Compared to the PNR case, the minima of the AMPNR surfaces gain rouhgly 5 MeV in binding energy and the absolute minimum is obtained for $J^\pi = 3/2^+$. It is also worth mentioning that the one-quasi-particle state giving the lowest projected state is obtained by blocking a quasi-particle that is dominated by a single-particle state originating from the spherical $2\text{d}_{3/2}$ shell.
The latter observations are consistent with the experimental spin-parity assignment $3/2_1^+$ for the ground state of \Au~as well as its naive single-particle interpretation. However, we notice that the minimum for the $J^\pi = 3/2^+$ surface is located at a deformation with an angle $\gamma=24^\circ$, which seems to be at variance with the oblate-like shape expected from simple arguments as mentioned above. 
Nevertheless, it is important to remark that the configuration mixing may change this picture. In addition, we displayed in Fig.~\ref{fig:au197_ener_ampnr_all} only the surface for the lowest  $J^\pi = 3/2^+$ projected states, but given the fact that we explore triaxial deformations, all the reference states with a non-zero average value of $\gamma$ will generate after angular-momentum restoration two projected states with $J^\pi = 3/2^+$ that will enter the configuration mixing.
Ultimately, given the fact that the AMPNR is only an intermediate step in our approach, it is neither possible nor desirable to definitively characterize the structure of the final correlated state at this level of approximation. 

Additionally, we note that the angular-momentum projection does not shift the energy minimum towards larger values of $\beta$ compared to the plain PNR case, which is contrary to what is often observed in MR-EDF calculations \cite{Bender08a,Rodriguez10a,Yao10a,Bally14a,Bally22a}.

\begin{figure}
    \centering
    \includegraphics[width=.80\linewidth]{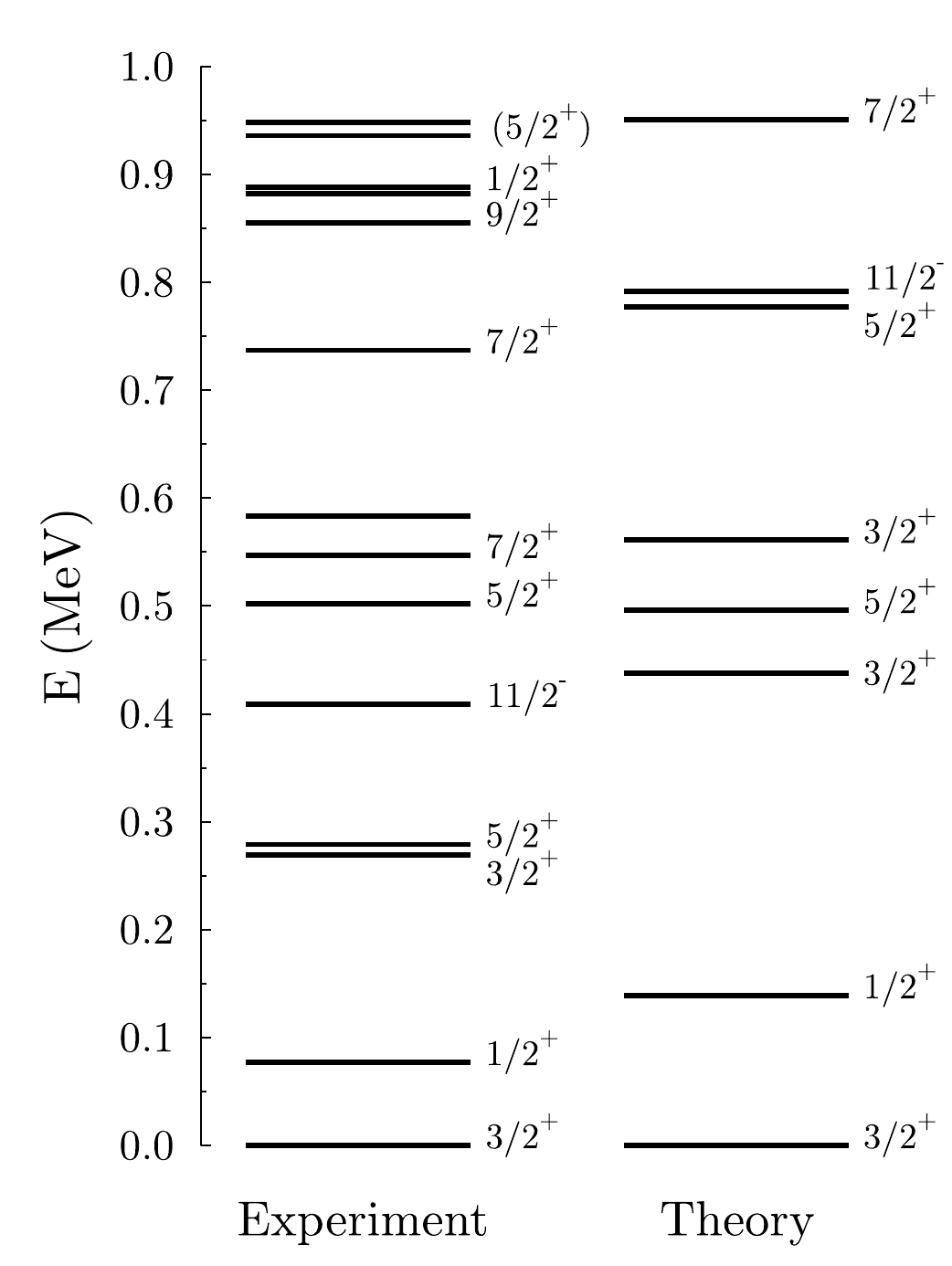}
    \caption{Low-energy spectrum for $^{197}$Au. Experimental data are taken from \cite{AIEAa}, which are based on the evaluation \cite{Huang05a}}
    \label{fig:au197_spec}
\end{figure}

\subsubsection{Low-energy spectroscopy}
\label{sec:spectrum}

Finally, we perform the full configuration mixing of symmetry projected reference states considering for positive (negative) parity a set containing 24 (19) one-quasi-particle states. In Fig.~\ref{fig:au197_spec} we compare the theoretical results to the available experimental data for the low-lying states up to 1 MeV of excitation energies. First of all, we remark that the theory is able to reproduce the spin-parity assignment for the ground state ($3/2_1^+$) as well as for the the first ($1/2_1^+$), second ($3/2_2^+$) and third ($5/2_1^+$) excited states. As in experimental data, the theory predicts a staggering between the two fomer and two latter states but the relative spacing between the two pairs of levels, as well as the spacing between the levels within a pair, are too large.
The $5/2_2^+$ and $7/2_1^+$ states also appear in our calculations but at too high excitation energy.

The low-lying spectrum of positive parity states of $^{197}$Au has been interpreted with De-Shalit's core-excitation model of odd-mass nuclei \cite{DeShalit61a}, within which an odd-even nucleus is treated as a single nucleon coupled to an even-even core. Whenever the excitation of the core is energetically favored compared to the promotion of the single nucleon to a higher orbital, in this model the lowest lying excited states of the odd-even nucleus can be interpreted as the single-particle configuration of the ground state coupled in different ways to the lowest excitation of the even-even core. When applied to $^{197}$Au \cite{Braunstein62a,DeShalit65a,McGowan71a,Powers74a,Bolotin79a}, the ground state of the nucleus is constructed as a proton $2\text{d}_{3/2}$ particle (hole) coupled to a $^{196}$Pt ($^{198}$Hg) core with $J^\pi=0^+$.\footnote{We mention in passing that some authors argue that using a $^{198}$Hg core provides a better global description of experimental data \cite{Bolotin79a}.} Then, the weak coupling of the same $2\text{d}_{3/2}$ particle, or hole, to the $J^\pi=2^+$ excited state of the (appropriate) core generates a quartet of state with $J^\pi = 1/2^+,~3/2^+,~5/2^+,~7/2^+$ whose energy centroid $E_c = \left\{ \sum_J (2J+1) E(J^\pi) \right\} / \left\{ \sum_J (2J+1) \right\}$ is equal to the energy $E(2^+)$ of the excited core \cite{Lawson57a}. Using the experimental excitation energies, we obtain $E_c = 364.2$ keV that is close to the values of $E(2^+) = 355.7$ keV and for $411.8$ keV for $^{196}$Pt and $^{198}$Hg, respectively. Computing the energy centroid within our approach, we obtain the value $E_c = 630.7$ keV that is obviously too large compared to the experimental one but should be compared the theoretical values of $E(2^+)$ for the neighboring even-even nuclei calculated within the same theoretical framework, which is technically possible but falls outside the scope of the present article.

The MR-EDF theory also correctly predicts the  $11/2_1^-$ state to be lowest state of negative parity but with an excitation energy of 792 keV, about 400 keV too high compared to the experimental value of 409 keV. This is especially surprising given that the energy difference between the AMPNR minima for $J^\pi = 3/2^+$ and $11/2^-$ has the correct order of magnitude as can be seen Figs.~\ref{fig:au197_ener_ampnr_all}. As a matter of fact, the minimum for $J^\pi = 11/2^-$ has roughly the same energy as the one for $J^\pi = 1/2^+$. What happens is that during the configuration mixing, the $11/2_1^-$ state does not gain nearly as much correlation energy as the positive parity states and, therefore, ends up at a too high excitation energy. It is not entirely clear why the mixing is less important in this case. It might be due to the deficiency of the effective interaction but we can not exclude the possibility that other factors may play a role. For example, we had to use more restrictive
values for the cutoffs before and after $K$-mixing to remove states not well represented on our Cartesian mesh. Therefore, it is possible that some important components or non-diagonal matrix elements suffer from numerical inaccuracy. Another possibility is that our selection strategy for the one-quasi-particle to self-consistent block at the mean-field level misses some configurations with negative parity relevant in the subsequent shape mixing.

In general, the theoretical spectrum is too spread in energy, which is an often-encountered deficiency of MR-EDF calculations based on reference states generated by a variation of the total energy without consideration for the angular momentum of the trial states. Indeed, such a variation tend to energetically favor the ground state. This deficiency can be in principle corrected by adding a constraint on the average angular momentum of the trial states during the minimization and using the value of the constraint as an additional generator coordinate. Unfortunately, such calculations are computationally expensive and very few practical applications exist \cite{Borrajo15a,Egido16b}.

\begin{table}
\centering
\begin{tabular}{ccc}
Quantity & Experiment & Theory \\
\hline
$E(3/2^+_1)$ & -1559.384 & -1556.044 \\[0.08cm]
$r_\text{rms}(3/2^+_1)$ & 5.4371(38) & 5.389 \\[0.08cm]
$\mu(1/2^+_1)$ & +0.416(3) & +0.01 \\[0.08cm]
$\mu(3/2^+_1)$ & +0.1452(2) & -0.38 \\[0.08cm]
$\mu(5/2^+_1)$ & +0.74(6) & +0.15 \\[0.08cm]
$\mu(5/2^+_2)$ & +3.0(5) & +0.14 \\[0.08cm]
$\mu(7/2^+_1)$ & +0.84(7) & +0.51\\[0.08cm]
$\mu(9/2^+_1)$ & +1.5(5) & +0.81 \\[0.08cm]
$\mu(11/2^-_1)$ & (+)5.96(9) & +6.87 \\[0.08cm]
$Q_s(3/2^+_1)$ & +0.547(16) & +0.65 \\[0.08cm]
$Q_s(11/2^-_1)$ & +1.68(5) &  +2.05 \\[0.08cm]
\hline
\end{tabular}
\caption{Spectroscopic quantities for the low-lying states of $^{197}$Au: total energy $E$ (MeV), root-mean-square (rms) charge radius $r_\text{rms}$ (fm), magnetic dipole moments $\mu$ ($\mu_N$), and spectroscopic quadrupole moments $Q_s$ ($e$b). Experimental data are taken from \cite{Angeli13a,Huang21a,Wang21a,Stone19a,Stone20a}. The experimental error on
the binding energy is much smaller than the rounded value given here}
\label{tab:spec197}
\end{table}

\begin{table}
\centering
\begin{tabular}{cccc}
Transition & Type & Experiment & Theory \\
\hline
$1/2^+_1 \rightarrow 3/2^+_1$ & E2 & 35(3) & 45 \\[0.08cm]
                              & M1 & 0.004 & 0.019 \\[0.08cm]
$3/2^+_2 \rightarrow 1/2^+_1$ & E2 & 18(3) & 6 \\[0.08cm]
                              & M1 & 0.089(9) & 0.048 \\[0.08cm]
$3/2^+_3 \rightarrow 1/2^+_1$ & E2 &  & 9 \\[0.08cm]
                              & M1 &  & 0.34 \\[0.08cm]
$3/2^+_2 \rightarrow 3/2^+_1$ & E2 & 18.5(19) & 0.4 \\[0.08cm]
                              & M1 & $<$ 0.001  &  0.002 \\[0.08cm]
$3/2^+_3 \rightarrow 3/2^+_1$ & E2 &  & 4 \\[0.08cm]
                              & M1 &  &  0.02 \\[0.08cm]
$5/2^+_1 \rightarrow 1/2^+_1$ & E2 & 14.4(17) & 12 \\[0.08cm]
$5/2^+_1 \rightarrow 3/2^+_1$ & E2 & 26(6) & 30 \\[0.08cm]
                              & M1 & 0.034(4) &  0.065\\[0.08cm]
$5/2^+_2 \rightarrow 1/2^+_1$ & E2 & 7.6(23) & 8 \\[0.08cm]
$5/2^+_2 \rightarrow 3/2^+_1$ & E2 & 7(6) & 0.4 \\[0.08cm]
                              & M1 & 0.083(10) & $<$ 0.001 \\[0.08cm]
$7/2^+_1 \rightarrow 5/2^+_1$ & E2 & 0.18(7) & 1 \\[0.08cm]
                              & M1 & 0.012(1) & 0.106 \\[0.08cm]
$7/2^+_1 \rightarrow 3/2^+_1$ & E2 & 33(3) & 38 \\[0.08cm]
$7/2^+_1 \rightarrow 3/2^+_2$ & E2 & 6.8(20) & 0.3 \\[0.08cm]
$7/2^+_1 \rightarrow 3/2^+_3$ & E2 &  & 3 \\[0.08cm]
$7/2^+_2 \rightarrow 3/2^+_2$ & E2 & 6(4) & 22 \\[0.08cm]
$7/2^+_2 \rightarrow 3/2^+_3$ & E2 &  & 2 \\[0.08cm]
$7/2^+_2 \rightarrow 5/2^+_1$ & E2 & 21(6) & 13 \\[0.08cm]
                              & M1 & 0.175(23) & 0.010 \\[0.08cm]
$9/2^+_1 \rightarrow 7/2^+_1$ & E2 & 10(7) & 10 \\[0.08cm]
                              & M1 & 0.028(10) & 0.047 \\[0.08cm]
$9/2^+_1 \rightarrow 5/2^+_1$ & E2 & 41(5) &  43 \\[0.08cm]
\hline
\end{tabular}
\caption{Reduced transition probabilities among the low-lying state of $^{197}$Au given in Weisskopf units. Experimental data are taken from \cite{AIEAa}, which are based on the evaluation \cite{Huang05a}}
\label{tab:trans197}
\end{table}

In Table \ref{tab:spec197}, we report spectroscopic quantities for some of the low-lying states. First, we see that the calculations reproduce fairly well the binding energy and root-mean-square charge radius of the ground state, with a relative accuracy below 1\%. The spectroscopic quadrupole moments for the $3/2_1^+$ and $11/2_1^-$ states are also reasonably well described in spite of being slightly too large. While we indicate in Table \ref{tab:spec197} the value for spectroscopic quadrupole moment of the ground state, $Q_s(3/2^+_1) = 0.547(16)$ $e$b, currently taken as the accepted value in the compilation of Ref.~\cite{Stone16a}, and which was determined using muonic hyperfine measurments \cite{Powers74a}, we remark that other values appear in the literature that are slightly larger, i.e.\ 0.60 $e$b and 0.64 $e$b in Ref.~\cite{Schwerdtfeger05a} and 0.59 $e$b in Ref.~\cite{Itano06a}, and in better agreement with the value of 0.65 $e$b obtained in our calculations.

Concerning the magnetic moments, they are, overall, poorly described. The values of most of them are significantly underestimated in our calculations and the moment of the ground state has even the wrong sign.
Surprisingly, the best (relative) agreement with experimental data is obtained for the magnetic moment of the $11/2_1^-$ state. 
A similar mediocre description of the magnetic moments was already observed in our study of the $^{128,129,130}$Xe nuclei and we refer to this article  \cite{Bally22b} for a discussion of the large spectrum of possible reasons for this problem that is faced by the vast majority of EDF calculations of magnetic properties.

In Table \ref{tab:trans197}, we compare the theoretical values for the reduced transition probabilities $B(E2)$ and $B(M1)$ to available experimental data. Concerning the $E2$ transitions, the theory gives reasonable estimates for most of the decays. In particular, all of the strong transitions, i.e.\ $1/2^+_1 \rightarrow 3/2^+_1$, $5/2^+_1 \rightarrow 3/2^+_1$, $7/2^+_1 \rightarrow 3/2^+_1$ and  $9/2^+_1 \rightarrow 5/2^+_1$, are well described. More generally, the hierarchy between the transitions seems to be respected, i.e.\ strong (weak) experimental transitions tend to be strong (weak) in our calculations. One notable exception are the transitions towards/from the $3/2_2^+$ state that are largely underestimated in our calculations. A possible interpretation is that the $3/2_2^+$ and $3/2_3^+$ states are inverted in our calculation compared to the experimental spectrum. Indeed, in Table \ref{tab:trans197} we also report the calculated transitions towards/from the $3/2_3^+$ state that are in better agreement with the experimental data for the transitions towards/from  $3/2_2^+$ state.
In the limit case of the core-excitation model discussed above, the reduced transition probabilities from the states of quartet $1/2_1^+,~3/2_2^+,~5/2_1^+,~7/2_1^+$ to the $3/2_1^+$ ground state are supposed to be equal among each other and with the $B(E2:2_1^+ \rightarrow 0_1^+)$ of the even-even core. Obviously, these equalities are not verified exactly for experimental data but the values remain somewhat close.\footnote{We mention that the $B(E2:2_1^+ \rightarrow 0_1^+)$ values are 40.6(20) and 28.8(4) W.u.\ for $^{196}$Pt and $^{198}$Hg, respectively \cite{AIEAa,Huang07a,Huang16a}.} In particular, within the same model, the electromagnetic transition probabilities are very sensitive to the mixing of the $J^\pi = 3/2^+$ intrinsic states \cite{Braunstein62a}, a problem that might also be present in our approach.   

Concerning the $M1$ transitions, the model performs poorly and most of the probabilities are either widely underestimated or widely overestimated. These lacking results are consistent with the observation made above on the magnetic moments. Again, this characteristic is a deficiency found in many nuclear EDF calculations. While the projection techniques used here are crucial for the reliable and unambiguous comparison of calculated and experimental data for magnetic properties, they do by themselves not lead to a satisfying description of data. We refer again to \cite{Bally22b} for further discussion of this issue.

\subsubsection{Collective wave functions}
\label{sec:collective}

We now turn our attention towards the analysis of the collective wave functions $g^{J^\pi}_{\sigma}(\beta,\gamma)$ of the correlated states as defined in Ref.~\cite{Bally22b}. We recall here that the squared collective wave functions (scwf) is a quantity that can be used to gauge the importance of a given deformation in the correlated wave function obtained in the final step of the MR-EDF calculations, with the caveat that, strictly speaking, it cannot be interpreted as a probability distribution due to the non-orthogonality of the reference states in the set.

\begin{figure*}[t]
\centering
\includegraphics[width=.245\linewidth]{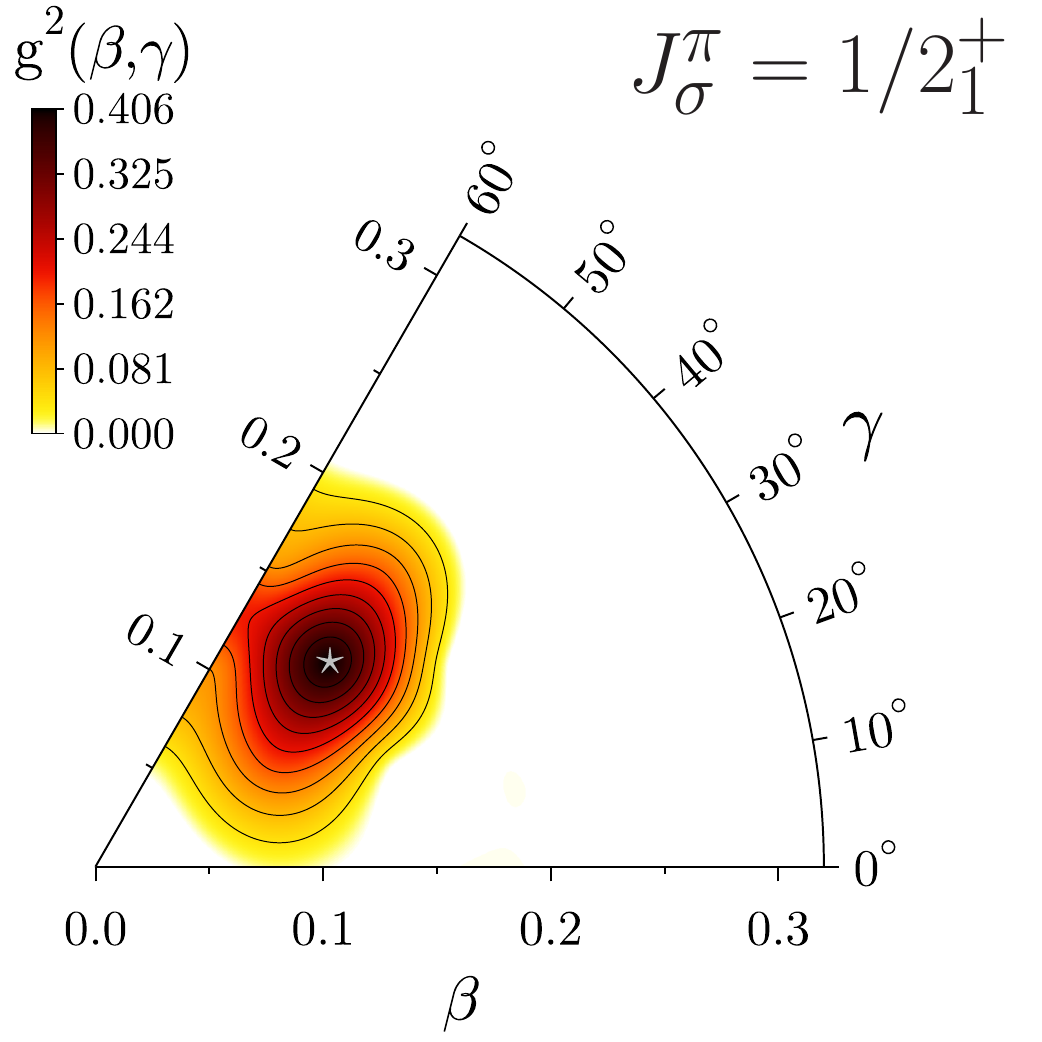}
\includegraphics[width=.245\linewidth]{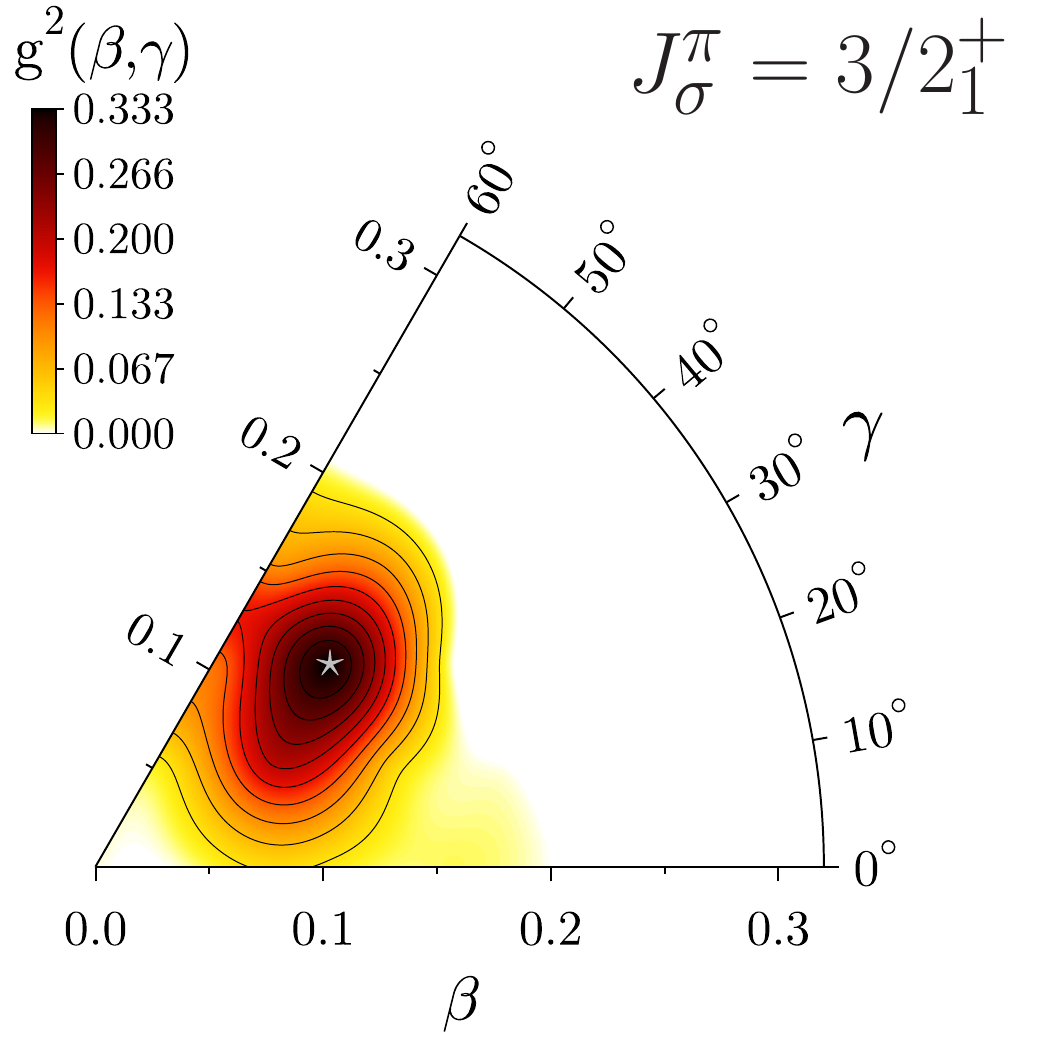} 
\includegraphics[width=.245\linewidth]{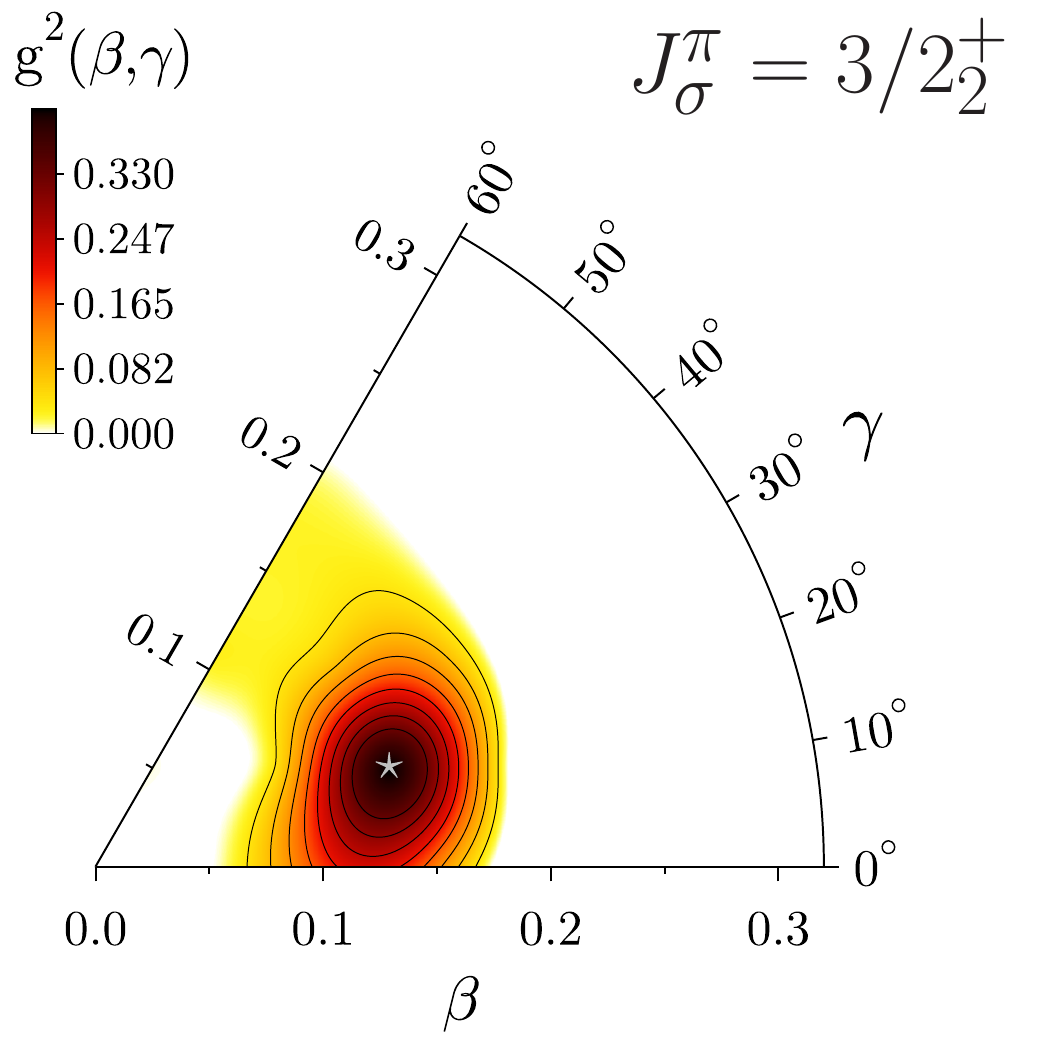}
\includegraphics[width=.245\linewidth]{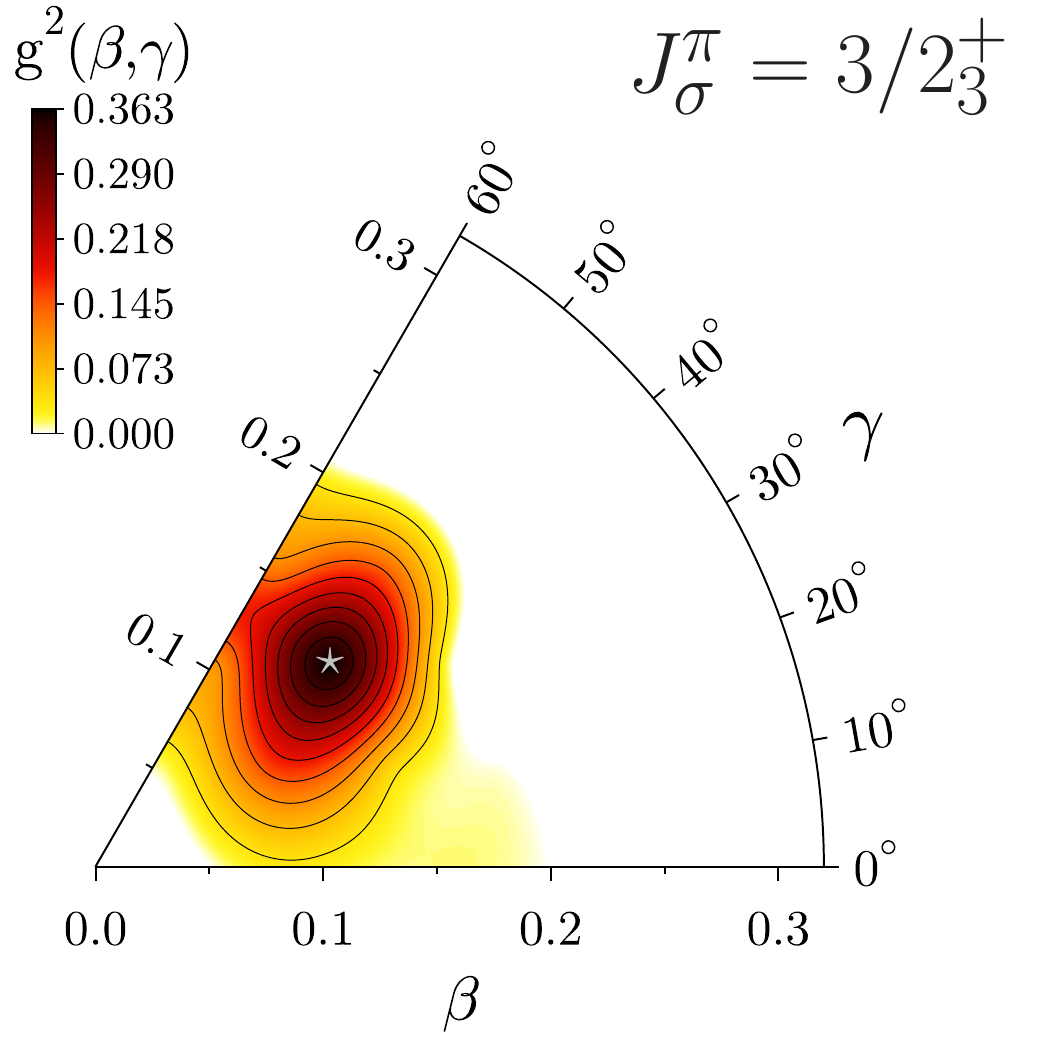} 
\includegraphics[width=.245\linewidth]{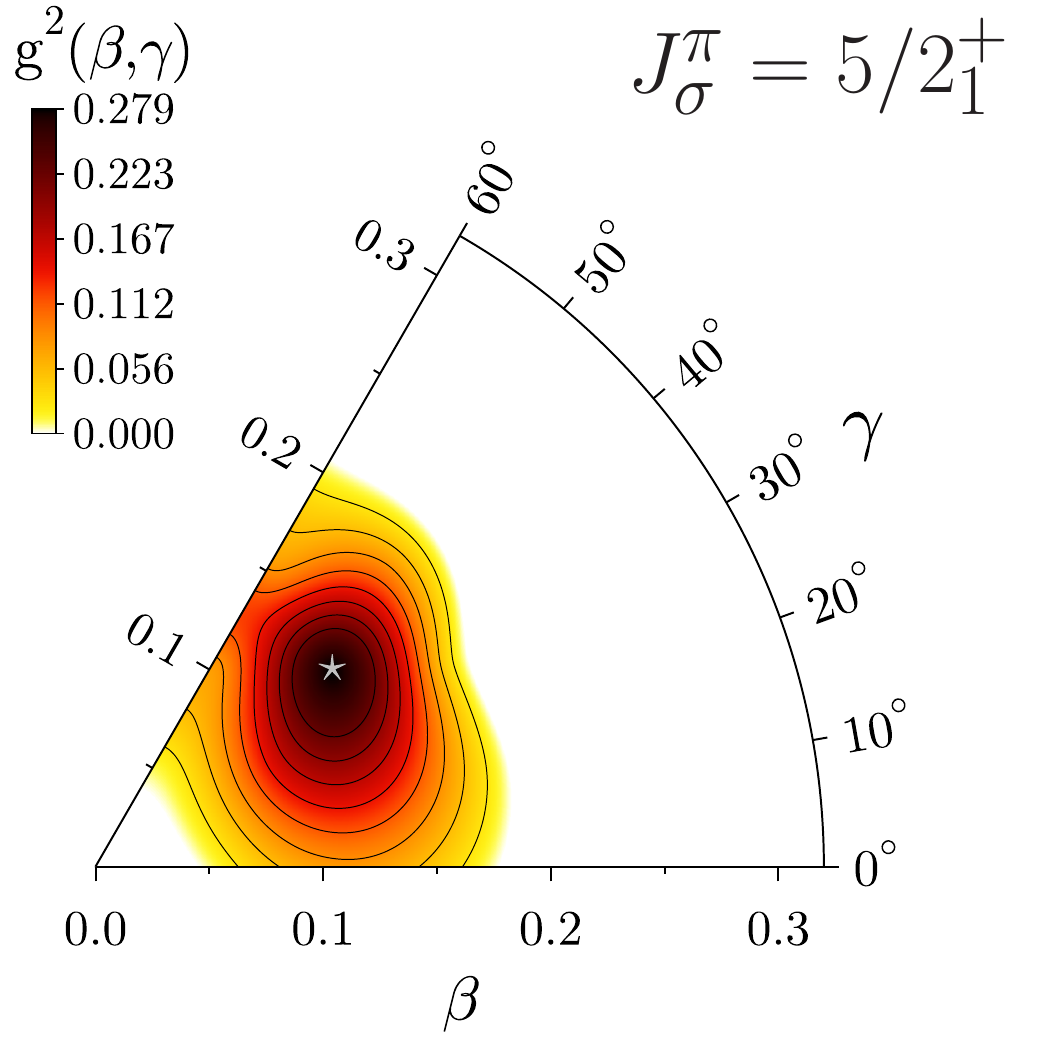}
\includegraphics[width=.245\linewidth]{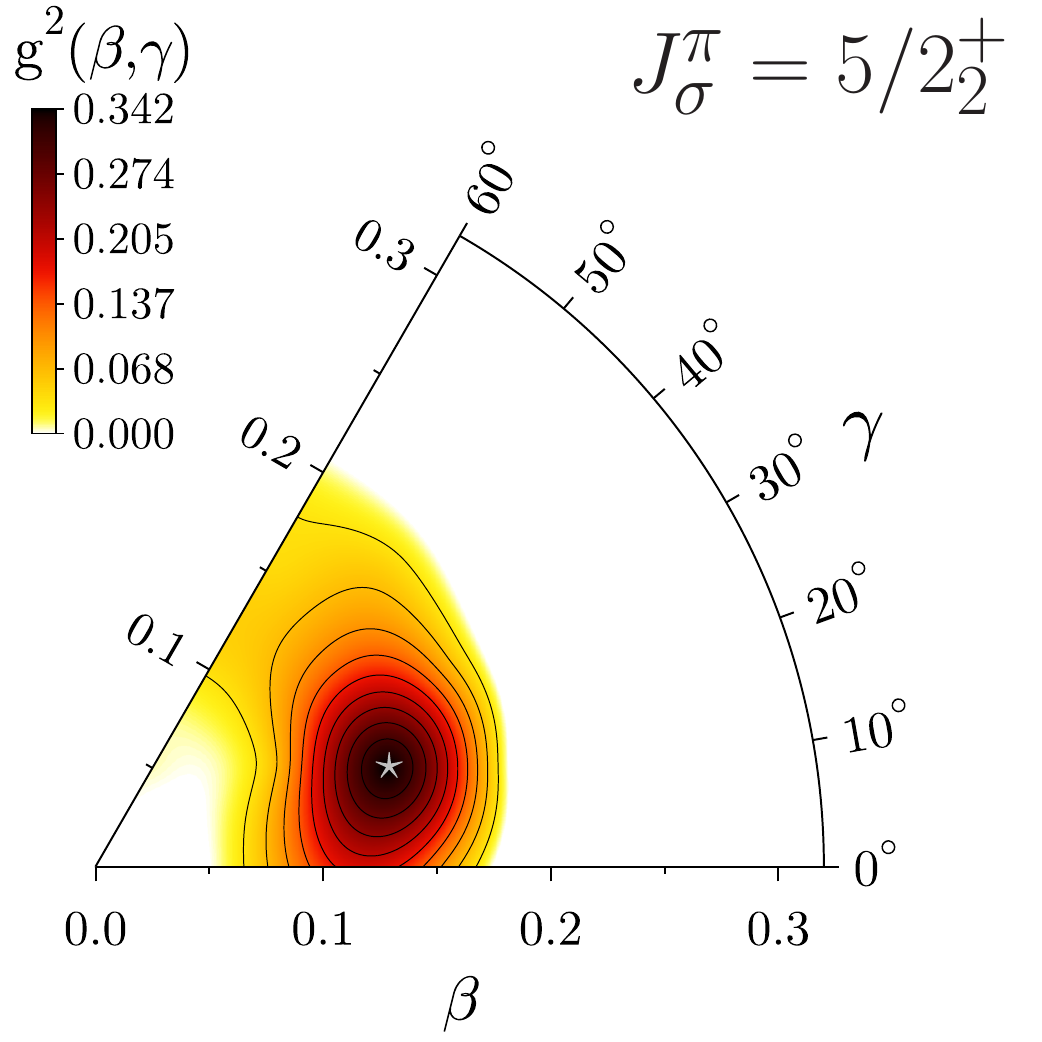}
\includegraphics[width=.245\linewidth]{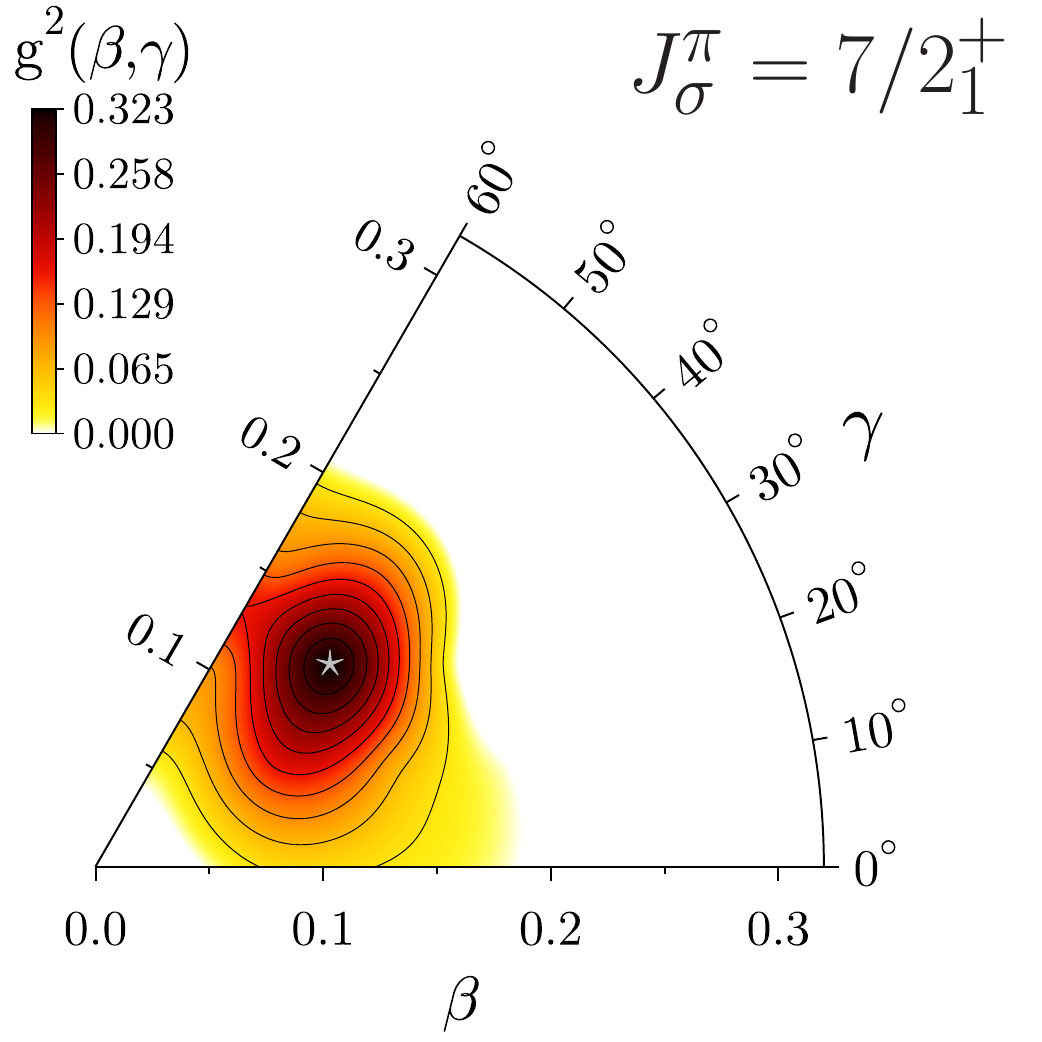}
\includegraphics[width=.245\linewidth]{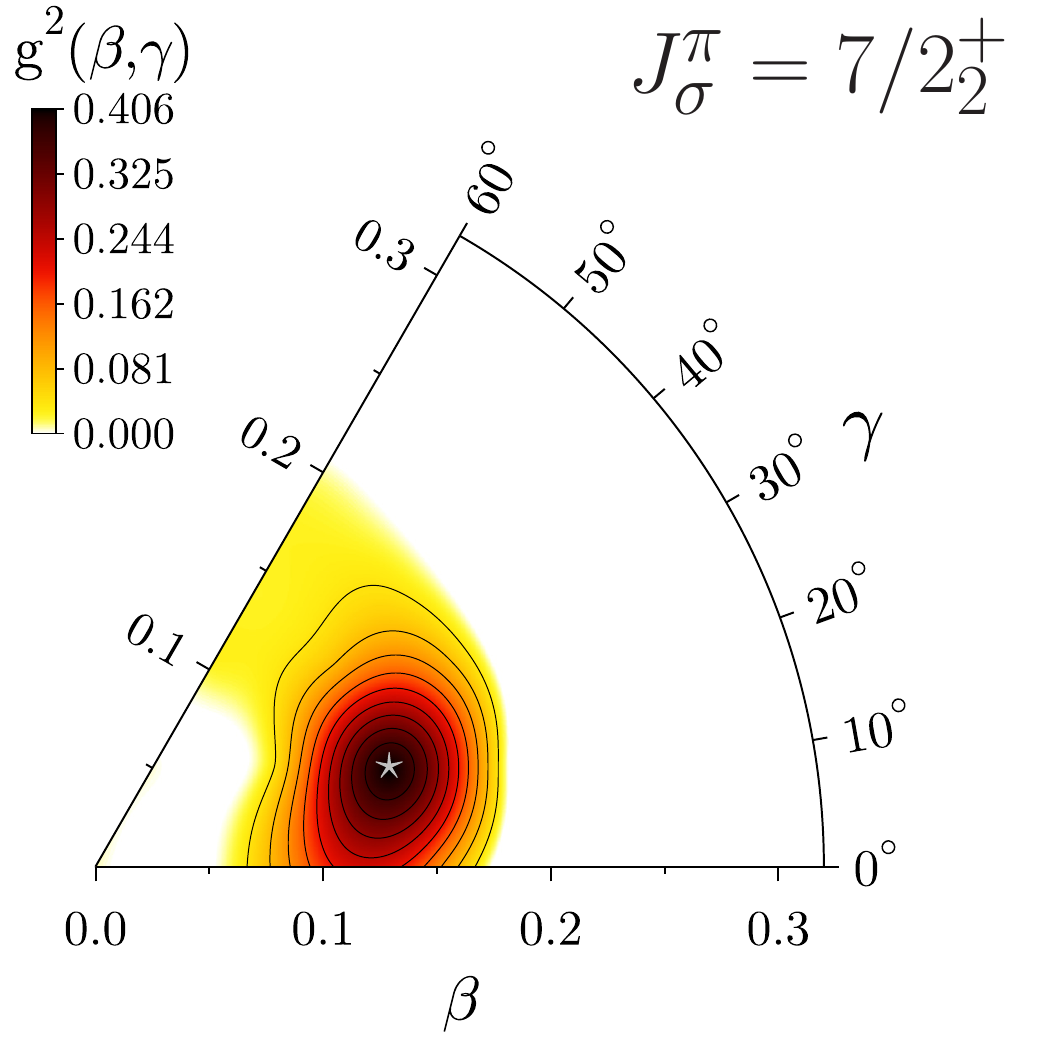}
\includegraphics[width=.245\linewidth]{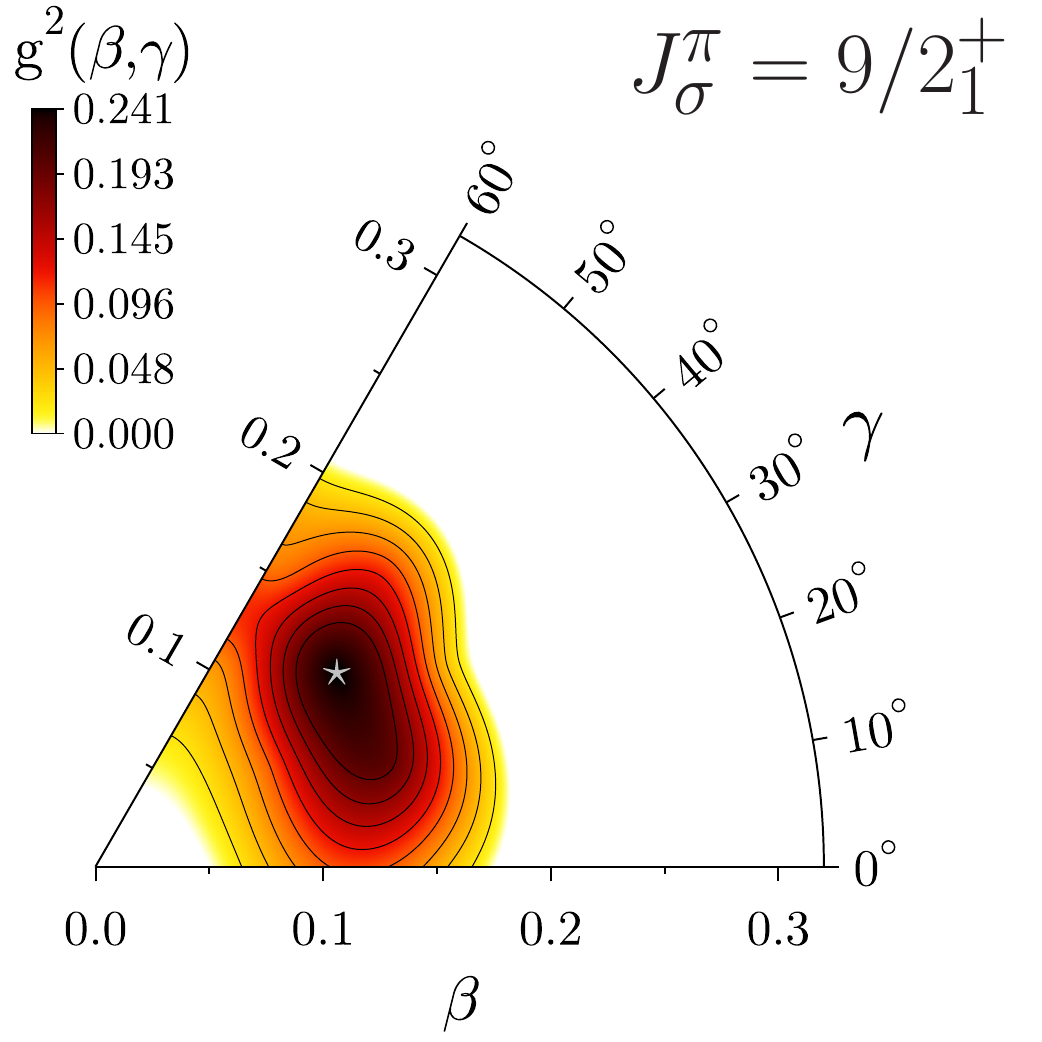}
\includegraphics[width=.245\linewidth]{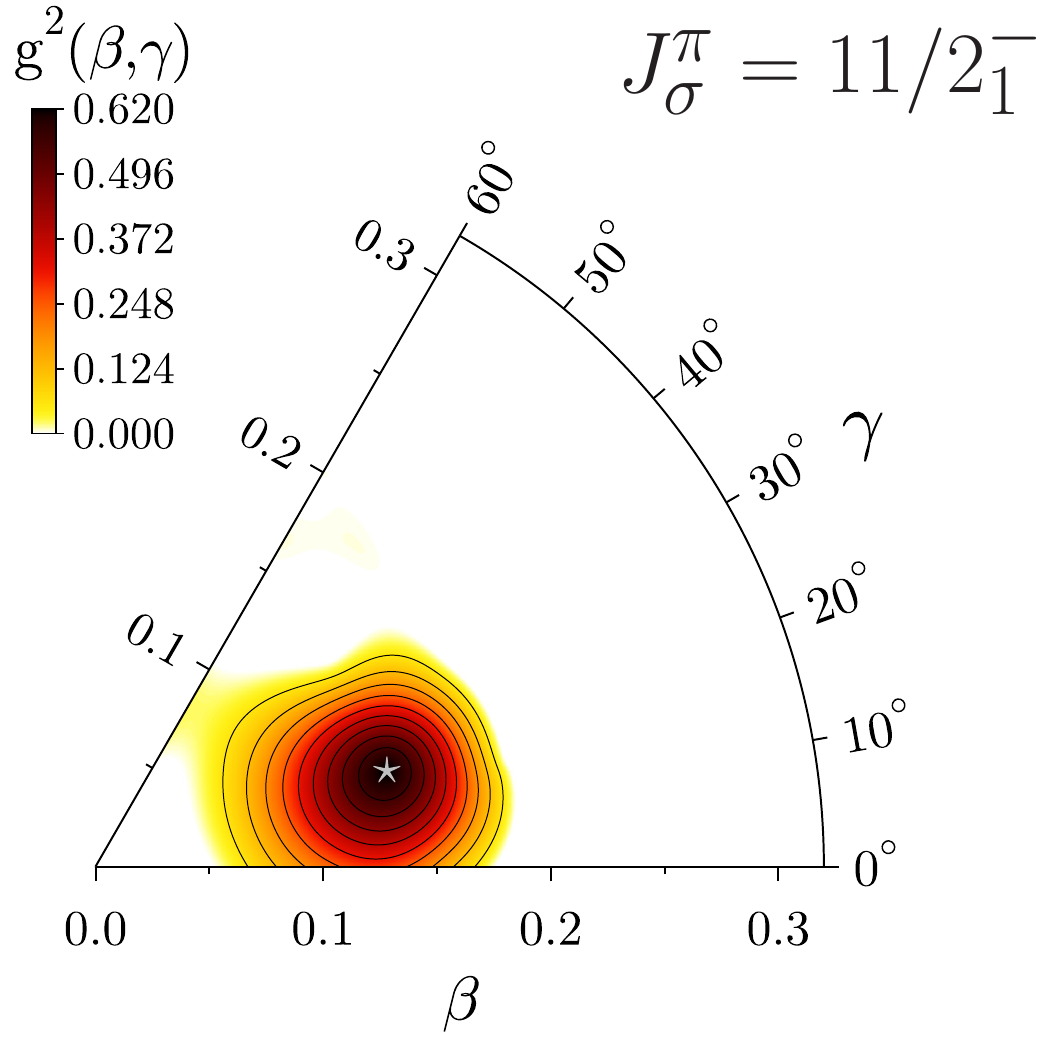}
\caption{Collective wave function squared for several low-lying states of $^{197}$Au with different values of $J^\pi_\sigma$. Black lines are separated by 10\% of the (respective) maximum value indicated by a silver star}
    \label{fig:au197_coll}
\end{figure*}

In Fig.~\ref{fig:au197_coll}, we display the scwf for several low-lying states of \Au. Interestingly, in all cases, the distribution of the scwf squared is dominated by triaxial shapes, with a sharply peaked maximum that has a quadrupole deformation of $\beta \simeq 0.14$. But depending on the value of $J^\pi_\sigma$, the maximum is either located at angle $\gamma \approx 40^\circ$ or $\gamma \approx 20^\circ$. In particular, we remark that the scwf of the $3/2_1^+$ and $3/2_2^+$ states exhibit different behaviour with the former being located closer to the oblate axis whereas the latter favours the prolate side of the sextant. Also, the scwf of the $3/2_3^+$ state is very similar to the one of the $3/2_1^+$ state. This is interesting for two reasons. First, it is contrary to what could have been expected looking at the AMPNR energy surface for $J^\pi = 3/2^+$ in Fig.~\ref{fig:au197_ener_ampnr_all}. Second, this is consistent with the oblate-like behavior expected in an independent-particle model for a nucleus close to the $Z=82$ and $N=126$ shell closures. Still, it is important to stress that non-axial deformations carry a substantial percentage of the scwf. 

Looking more closely at the scwfs of the positive parity states, we can arrange them into three groups of similar 
appearance: a) the states $1/2_1^+$, $3/2_1^+$, $3/2_3^+$ and $7/2_1^+$ that have a scwf mostly located in the range $30^\circ \le \gamma \le 60^\circ$ b) the states $3/2_2^+$, $5/2_2^+$ and $7/2_2^+$
that have a scwf mostly located in the range $0^\circ \le \gamma \le 30^\circ$ and c) the states $5/2_1^+$ and $9/2_1^+$ whose scwf are more evenly distributed as a function of $\gamma$ and seem a combination of the cases a) and b). This is in good agreement with the data for the reduced transition probabilities given in Table \ref{tab:trans197}. Indeed, the transitions between the states of a given group have very large $B(E2)$ values whereas the transitions between the states belonging to a different group are less likely. This is not a perfect rule, however, because the $5/2_1^+$ state has also an strong transition towards the $3/2_1^+$ ground state but not towards the $1/2_1^+$ excited state even if the scwfs of the two latter states have similarities. To come back to the core-excitation model analysis, the fact that the scwf of the $1/2_1^+$, $3/2_3^+$,\footnote{Provided that we interpret the $3/2_2^+$ and $3/2_3^+$ states as being inverted in our calculations compared to experimental data.}  $5/2_1^+$ and $7/2_1^+$ excited states have a large overlap with the scwf of the $3/2_1^+$ ground state is consistent with the interpretation of the quartet of positive parity state as being weak coupling of the same single-particle state to a collective even-even core with an angular momentum of either $J^\pi=0^+$ or $2^+$. 

As a last comment, we remark that the scwf of the $11/2_1^-$ state has a narrower distribution than the other ones displayed, which is consistent with the fact that this state does not mix as much when diagonalizing the Hamiltonian within the space spanned by the symmetry-projected reference states.

\subsubsection{Average deformation}
\label{sec:averagedef}

Finally, following the strategy presented in our previous article on xenon isotopes \cite{Bally22b}, we use the scwf to compute deformation parameters for the $3/2_1^+$ ground state of \Au~and obtain: an average elongation of $\bar{\beta}(3/2_1^+)=0.13$, with a standard deviation of $\Delta \beta (3/2_1^+)=0.03$, and an average angle of $\bar{\gamma}(3/2_1^+)=40^\circ$, with a standard deviation of $\Delta \gamma (3/2_1^+)=15^\circ$. This average deformation is consistent with the distribution displayed in Fig.~\ref{fig:au197_coll} as the maximum is located at a deformation of $\beta=0.14$ and $\gamma=41^\circ$ but the distribution extends towards smaller values of $\beta$ and is more or less equally distributed with respect to the $\gamma=40^\circ$ axis.

Within the rigid rotor model, it is also possible to compute a deformation $\beta_r$ for the $0^+_1$ ground state of an even-even nucleus using the experimental $B(E2)$ values, for more details see for example Refs.~\cite{RS80a,Bally22b}. Computing $\beta_r$ for the even-even nuclei adjacent to \Au~one obtains the value 0.13 for $^{196}$Pt and 0.11 $^{198}$Hg. Our average deformation  $\bar{\beta}(3/2_1^+)=0.12$ fits nicely between these two values, although we have to mention that the definitions of the two elongations are model dependent such that this excellent agreement may be partly accidental.

The results of axially-symmetric EDF calculations based on the Gogny D1S parametrization \cite{Decharge80a,Berger91a} reported in the AMEDEE database \cite{Hilaire07a} indicate a sharp minimum at a deformation of about $\beta \approx 0.12$ for \Au. This is perfectly consistent with our estimate. The AMEDEE database also reports average deformations obtained from large-scale five-dimensional collective Hamiltonian (5DCH) calculations of even-even nuclei throughout the nuclear chart \cite{Delaroche10a}. We recall here that the 5DCH can be derived as an approximation to the full GCM performed here \cite{RS80a}. While their definition for the average deformation differs from ours, we mention that for $^{196}$Pt ($^{198}$Hg), they obtain an average elongation of 0.135 (0.110), with a standard deviation of 0.032 (0.030), and average angle of $32^\circ$ ($31^\circ$), with a standard deviation of $12^\circ$ ($12^\circ$).
If the values for the elongation are consistent with our result, the average angles differ slightly with the 5DCH result indicating a deformation right at the center of the triaxial plane, although the fluctuations are large enough such that the results are compatible.

%
%
\section{Heavy-ion collisions}
\label{sec:collisions}

As previously mentioned, knowing the structure of $^{197}$Au is of particular relevance in the context of high-energy nuclear experiments, as gold is the primary species collided at the BNL RHIC. This section analyzes the consequences of our results for model simulations of ultrarelativistic $^{197}$Au+$^{197}$Au collisions. 

\subsection{Woods-Saxon parameterization of the ground state}

Traditionally, simulations of high-energy nuclear collisions take as input from nuclear structure a point-nucleon density which is used to sample nucleon coordinates and define an interaction region between two ions on a collision-by-collision basis.\footnote{More sophisticated calculations based on nuclear configurations obtained from \emph{ab initio} nuclear theory have also been recently performed \cite{Lim19a,Rybczy19a,Summerfield21a,Nijs22a}. For the moment, they are limited to the description of collisions of $^{16}$O ions.} The standard choice for the nucleon density is that of a deformed Woods-Saxon (WS) profile:
\begin{equation}
\label{eq:ws}
    \rho (r,\theta,\phi) = \frac{\rho_0}{1+e^{[r-R(\theta,\phi)]/a}},
\end{equation}
where $r,\theta,\phi$ are the usual spherical coordinates, $\rho_0$ is the saturation density, $a$ is the surface diffuseness and $R(\theta,\phi)$ is the nuclear radius parameterized as
\begin{align}
\label{eq:surf}
  R(\theta,\phi) &= 
  R_0 \biggl \{ 1 + \beta_{2}^{\rm WS} \biggl[ \cos(\gamma^{\rm WS}) Y_{20}(\theta,\phi) \\ &~+ \sqrt{2} \sin(\gamma^{\rm WS}) {\rm Re}\bigl (Y_{22}(\theta,\phi)\bigr)  \biggr ] + \beta_{4}^{\rm WS} Y_{40}(\theta,\phi) \biggr\}, \nonumber  
\end{align}
where the spherical harmonics $Y_{lm}(\theta,\phi)$ are in complex form. Note that the shape parameters $\beta_{2}^{\rm WS}$, $\gamma^{\rm WS}$ and $\beta_{4}^{\rm WS}$ represent surface deformations that differ from the volume deformation reported in the analysis of the previous sections \cite{Wouter}.

\begin{figure}[t]
    \centering
    \includegraphics[width=0.80\linewidth]{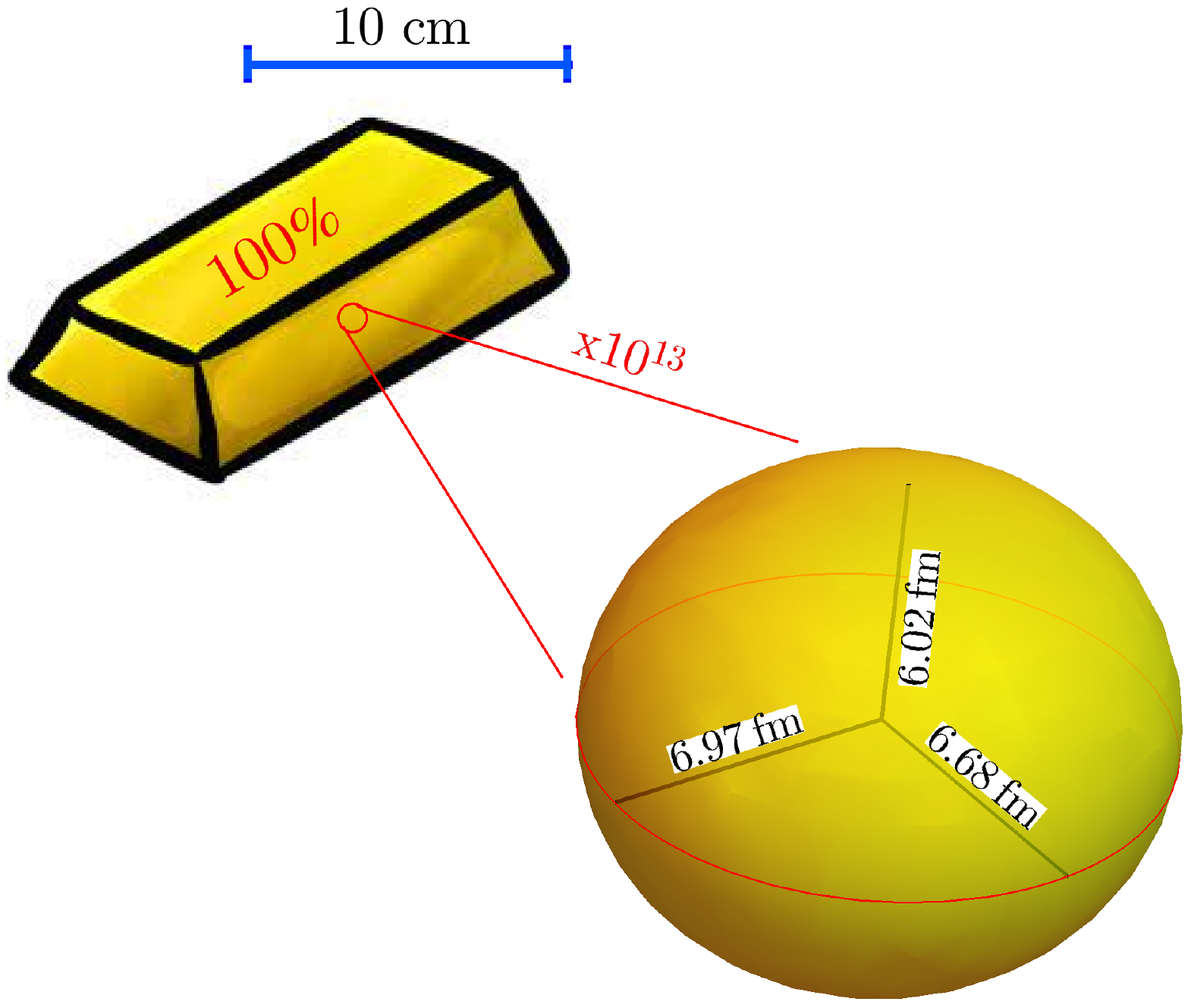}
    \caption{Schematic illustration of the shape of $^{197}$Au based on the surface parametrization of the matter density of Eq.~(\ref{eq:surf}), and using the parameters reported in Tab.~\ref{tab:1}
    }
    \label{fig:G2}
\end{figure}

We consider now the intrinsic shape of \Au{} computed from a single Hartree-Fock-Bogoliubov (HFB) calculation with the SLyMR1 interaction in which the expectation value of the quadrupole operators are constrained such that the one-body density of the trial one-quasi-particle state\footnote{The trial one-quasi-particle state is built by blocking a single-particle state originating from the spherical $2\text{d}_{3/2}$ shell.} verifies, on average, $\beta=\bar{\beta}(3/2_1^+)=0.13$ and $\gamma=\bar{\gamma}(3/2_1^+)=40^\circ$.\footnote{All other non-vanishing multipole moments authorized by the symmetries of our calculations are let free to adopt a value that minimizes the total energy of the trial quasi-particle state.}
We fit the resulting one-body nucleon density with the Woods-Saxon profile given in Eq.~(\ref{eq:ws}).  The fit parameters are reported in Tab.~\ref{tab:1}. We obtain, thus, a new microscopically motivated parametrization for the Woods-Saxon profile representing the nucleon density of the ground state of \Au{} which can be employed in simulations of high-energy collisions. This profile corresponds to a triaxial ellipsoid with radii 6.02 fm, 6.68 fm, and 6.97 fm, as illustrated in Fig.~\ref{fig:G2}.

\begin{table}[t]
\centering
\normalsize
\begin{tabular}{cccc}
Parameter & Proton &  Neutron  & Nucleon \\
\hline 
$\rho_0$  & 0.067 &  0.090 & 0.157 \\
$R_0$  &  6.44 &  6.65  &  6.56 \\
$a$ &  0.46 &  0.49  &  0.48 \\
$\beta_{2}^{\rm WS}$   &  0.134 &  0.137 & 0.135 \\
$\gamma^{\rm WS}$  & 43$^\circ$ &  43$^\circ$ &  43$^\circ$ \\
$\beta_{4}^{\rm WS}$   & -0.024 &  -0.023 &  -0.023 \\
\hline
\end{tabular}
\caption{\label{tab:1} 
Parameters for the point-proton, point-neutron and point-nucleon densities defined as in Eq.~\eqref{eq:ws} and fitted to reproduce the one-body densities of a quasi-particle state constrained to have, on average, $\beta=0.13$ and $\gamma=40^\circ$; see the body of the text for more details. The parameters $R_0$ and $a$ are given in units of fm, whereas $\rho_0$ is given in units of fm$^{-3}$
}
\end{table}

For completeness, we evaluate as well the neutron skin of the intrinsic shape, as defined by the difference of rms radii, $\Delta r_{np} = \langle r^2 \rangle_{n}^{1/2} - \langle r^2 \rangle_{p}^{1/2}$. For the density returned by the constrained HFB calculation, we find 
\begin{equation}
    \Delta r_{np} {\rm [HFB(\bar{\beta},\bar{\gamma})]} =  0.17 ~{\rm fm},
\end{equation}
which is in perfect agreement with the result obtained from the full MREDF calculation
\begin{equation}
    \Delta r_{np} {\rm [MREDF]} =  0.17 ~{\rm fm}.
\end{equation}
On the other hand, the fitted Woods-Saxon profile gives a neutron skin
\begin{equation}
    \Delta r_{np} {\rm [WS~fit]} = 0.19 ~{\rm fm},
\end{equation}
meaning that, even for a large nucleus such as $^{197}$Au, the Woods-Saxon parametrization does not fully capture skin differences of order 0.1 fm between neutrons and protons. We note that both the above estimates agree with a recent measurement of the STAR collaboration obtained via diffractive photo-production of $\rho^0$ mesons in ultra-peripheral $^{197}$Au+$^{197}$Au collisions \cite{Abdallah22a},
\begin{equation}
    \Delta r_{np} {\rm [STAR]} = 0.17 \pm 0.03~{\rm (stat.)} \pm 0.08~{\rm (syst.)} ~{\rm fm}. 
\end{equation}
We note, in addition, that the half-width radius obtained for $^{197}$Au by the STAR collaboration, $R_0 {\rm [STAR]}=6.53 \pm 0.06$ fm, is fully consistent with that exhibited by our nucleon density, $R_0 {\rm [WS~fit]}=6.56$ fm. This suggests that the density of gluons relevant for scattering at these beam energies is in fact very close to the rest-frame point-nucleon density. This potentially adds to the circumstantial evidence of a small nucleon width in high-energy collisions mediated by gluons \cite{Mantysaari16a,Giacalone22a,Mantysaari22a,Nijs22b,Giacalone22b}.

We discuss now the observational consequences of our newly-derived nucleon density for relativistic $^{197}$Au+$^{197}$Au collisions. Model calculations of such processes (see e.g. Ref.~\cite{Everett21a} for a state-of-the-art Bayesian analysis) have so far employed the charge density of the nucleus, as inferred from low-energy electron-nucleus scattering experiments \cite{DeVries87a}, as a proxy for the nucleon density. The corresponding radial profiles are $R_0=6.38$ fm, and $a=0.53$ fm. Nuclear quadrupole deformation has been instead included by simply implementing $\beta_2^{\rm WS}=-0.13$, as reported by finite-range liquid drop model evaluations \cite{Moller16a}. In terms of radial profiles, there are, thus, minor differences between the WS parametrization that we show in Tab.~\ref{tab:1} and that implemented in the literature. We only note a reduction by 0.05 fm in the diffuseness parameter, $a$, which is due to the inclusion of the neutron density. This will have a mild, though visible impact on the initial eccentricities, $\varepsilon_n$, of the system \cite{Shou15a,Xu21a,Jia21a}. A new feature of our calculation is instead the fact that $^{197}$Au is not fully oblate, but presents $\gamma^{\rm WS}=43^\circ$. We investigate now the impact of such a feature on high-energy collisions.

\subsection{Impact of the triaxiality}
In the context of multi-particle correlation measurements in the soft sector of high-energy nuclear collisions, the strongest sensitivity to the triaxial structure of the colliding nuclei is carried by the mean momentum-elliptic flow correlation \cite{Giacalone20a,Giacalone:2020awm,Jia22a},
\begin{equation}
\label{eq:PC}
    \rho_2 \equiv \rho(\langle p_t \rangle, v_2^2) = \frac{\langle \langle p_t \rangle  v_2^2 \rangle  - \langle \langle p_t \rangle  \rangle \langle v_2^2 \rangle}{\sigma(\langle p_t \rangle) \sigma( \langle v_2^2 \rangle)},
\end{equation}
where outer brackets denote a statistical average over events, and $\sigma(o)$ is the standard deviation of observable $o$. This quantity can be evaluated in the final states as a three-particle correlation \cite{Bozek16a}, and it measures the strength of the statistical correlation between the charged-particle average transverse momentum, $\langle p_t \rangle$, and the charged-particle elliptic flow, $v_2$, at a given collision multiplicity. 

To assess the impact of $\gamma^{\text{WS}}=43^\circ$ on the $\rho_2$ correlator of $^{197}$Au+$^{197}$Au collisions, we follow Ref.~\cite{Bally22a} and provide an estimate of the measured $\rho_2$ from high-statistics simulations of the initial condition of these processes. For the details of such simulations, we refer to the exhaustive descriptions given in Ref.~\cite{Bally22a}. Briefly, we assume that the distribution of final-state multiplicities is proportional to the distribution of initial-state entropy, $S$, which we calculate event-to-event following the original \trento{} parametrization \cite{Moreland15a} ($s ({\bf x}, \tau_0) \propto \sqrt{T_AT_B}$, $S =\int d^2{\bf x} ~s$) with a nucleon size $w=0.5$ fm, and a fluctuation parameter, $k$, tuned to reproduce measured multiplicity histograms in $^{208}$Pb+$^{208}$Pb collisions at CERN LHC energy. We consider that i) the mean transverse momentum is, at a given entropy, proportional to the initial $E/S$, where $E$ is the total energy of the system \cite{Gardim21a,Giacalone21c}, obtained upon application of the equation of state of high-temperature QCD ($e ({\bf x}) \propto s({\bf x})^{4/3}$, $E=\int d^2{\bf x}~ e$), and ii) that the elliptic flow is proportional to the initial eccentricity of the system, $\varepsilon_2$. The Pearson correlation coefficient of Eq.~(\ref{eq:PC}) can then be estimated by replacing $v_2^2$ and $\langle p_t \rangle$ with, respectively, $\varepsilon_2^2$ and $E/S$. Note that the resulting estimator should not be compared directly to the experimental measurements, as it misses effects related to the cuts in transverse momentum, $p_t$, implemented in the experimental analysis, which have been shown to be sizable for the magnitude of this observable \cite{Aad19a,Acharya22a,ATLAS:2022dov}. That said, it is the initial-state estimator that carries the dependence on the deformation parameters, such that the relative impact of the value of $\gamma^{\rm WS}$ on the final-state result can be assessed from it \cite{Bally22a,Jia:2021qyu}.

\begin{figure}
    \centering
    \includegraphics[width=\linewidth]{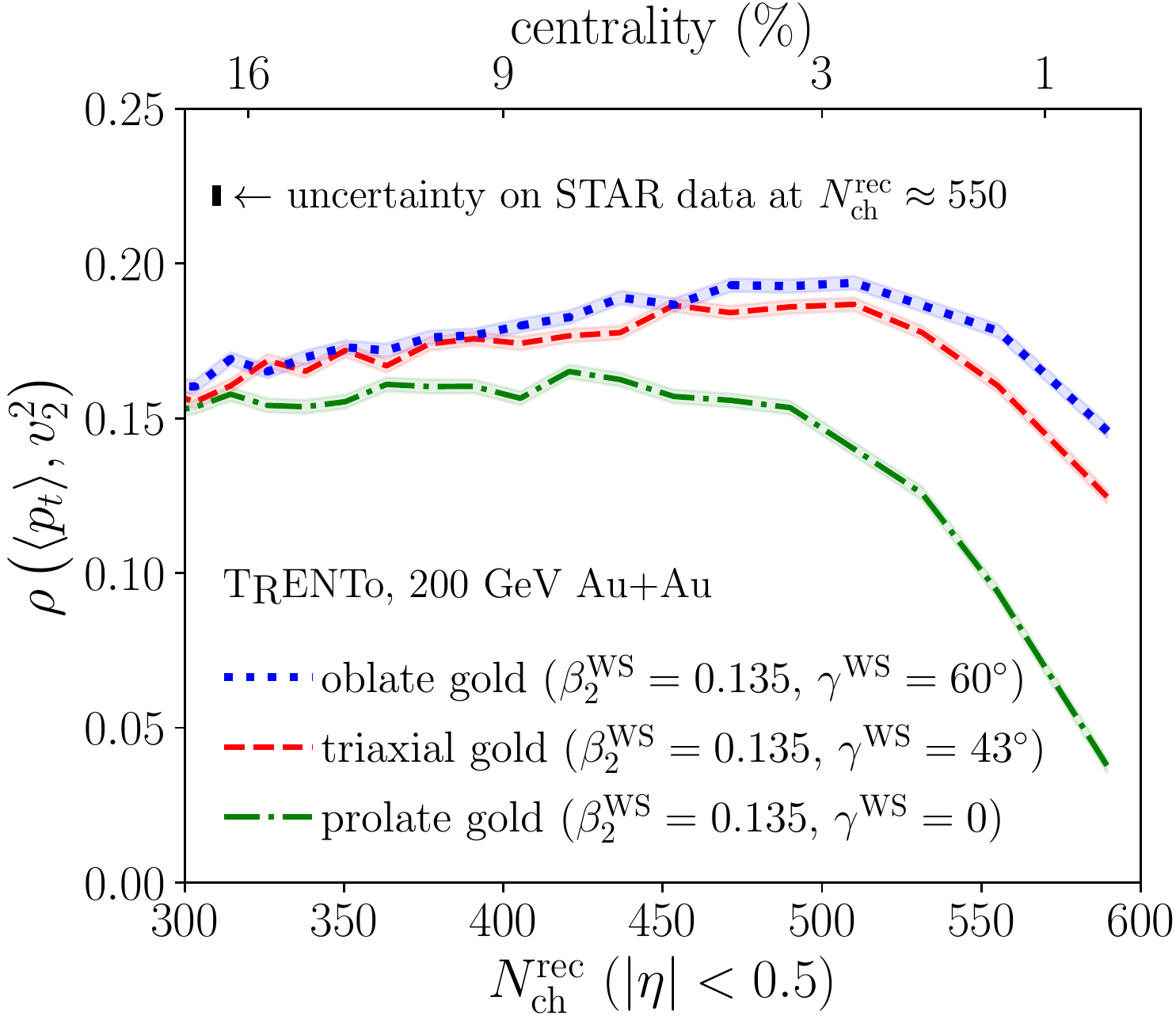}
    \caption{Initial-state estimates of $\rho(\langle p_t \rangle, v_2^2)$ in 200 GeV $^{197}$Au+$^{197}$Au collisions for prolate ions (dot-dashed line), oblate ions (dotted line) and triaxial ions (dashed line) presenting $\gamma^{\text{WS}}=43^\circ$, as a function of the number of reconstructed charged tracks in the STAR detector.  Shaded bands (of the same width as the lines) are statistical uncertainties. The figure reports as well the total uncertainty on preliminary STAR measurements for this observable at high multiplicities. }
    \label{fig:G1}
\end{figure}

We perform $20\times10^6$ minimum bias simulations of $^{197}$Au+$^{197}$Au collisions for three structure scenarios, namely, we set $\beta_2^{\rm WS}=0.135$, and consider $\gamma^{\rm WS}=0^\circ$, $43^\circ$, and $60^\circ$.\footnote{We safely neglect the effect of the very small hexadecapolarity of the nucleus, $\beta_4^{\rm WS}=-0.023$, in these simulations.} Rescaling the \trento{} entropy to match the observed mutliplicity of reconstructed charged tracks in the STAR detector, $N_{\rm ch}^{\rm rec}$, at midrapidity ($|\eta|<0.5$), our results for $\rho_2$ are reported in Fig.~\ref{fig:G1}. Qualitatively, the impact of $\gamma^{\rm WS}$ follows the generic parametric expectation $\rho_2 \propto c_0 - c_1 (\beta_2^{\rm WS})^3 \cos(3\gamma^{\rm WS})$, where $c_0$ and $c_1$ are positive coefficients \cite{Bally22a,Jia:2021qyu}. We conclude that a $17^\circ$ deviation from oblateness in $^{197}$Au leads to a correction of order 10-15\% to $\rho_2$ for collisions in the 0-2\% centrality range. We reiterate that, while our results for the magnitude of 
the Pearson coefficient should not be compared directly to data, we expect the correction induced by the triaxiality, relative to the oblate scenario, to be robustly captured by our initial-state evaluation. In Fig.~\ref{fig:G1} we report as well the size of the experimental error on preliminary $\rho_2$ data at high multiplicity from the STAR collaboration \cite{Zhang:2022sgk}. The error bar turns out to be significantly smaller than the splitting that we find between the triaxial scenario (red dashed line) and the oblate scenario (dotted blue line). Therefore, according to our results the impact of the triaxiality has been already isolated in the preliminary data, and it will be possible to quantify it in the future via high-precision hydrodynamic simulations. We stress, though, that the most effective way to access the value of $\gamma^{\rm WS}$ is by studying the $\rho_2$ correlator of $^{197}$Au+$^{197}$Au collisions normalized with that of $^{238}$U+$^{238}$U collisions, as done in Refs.~\cite{Bally22a,ATLAS:2022dov} to extract such an information in the comparisons of $^{129}$Xe+$^{129}$Xe and $^{208}$Pb+$^{208}$Pb collisions, which allows one to fully cancel theoretical and experimental systematical uncertainties and isolate transparent information about the nuclear structure. The current mismatch between hydrodynamic results and experimental data for $^{238}$U+$^{238}$U collisions \cite{Giacalone21b} prevents us, for the moment, from performing such an analysis, which will be thus reported in future work.

%
%
\section{Conclusions}
 \label{sec:conclu}
 
In the present article, we first reported on new results on the low-energy structure of the heavy odd-mass nucleus \Au~obtained by performing state-of-the-art MR-EDF calculations that include 
the mixing of angular-momentum and particle-number projected Bogoliubov quasi-particle states with different average triaxial shapes. All the calculations were realized using the parametrization SLyMR1 of a Skyrme-type pseudo-potential \cite{Sadoudi13a,JodonPHD}. 

Although odd-mass nuclei represent half of the existing nuclei in the nuclear chart, their calculations within the full-fledged MR-EDF framework are still scarce, exceptions being \cite{Bally14a,Borrajo17a,Borrajo18a,Bally22b}. In this work, to generate reference states adapted to the modeling of odd-mass nuclei, we performed self-consistent blocking of Bogoliubov one-quasi-particle states and considered exactly all the time-odd terms of the functional.

The results obtained on the low-energy spectroscopy of \Au~are reasonable. The spin-parity assignments for the $3/2^+_1$ ground state and for the first few excited states are correct even if the levels are too spread out, a well-known deficiency of usual MR-EDF calculations that can be corrected by adding a supplemental constraint on the average angular momentum of the trial wave functions when generating the set of reference states to be projected and mixed \cite{Borrajo15a,Egido16b}. The binding energy, root-mean-square charge radius and  spectroscopic quadrupole moment of the of the ground state are also well reproduced. By contrast, the calculations fail to reproduce the known magnetic moments for the ground and excited states.
Concerning the electromagnetic transitions, the values for the reduced transition probabilities $B(E2)$ are, overall, well described whereas the values for the $B(M1)$ are off, sometimes by more than one order of magnitude.

Starting from the collective wave function of the ground state, we computed average triaxial deformation parameters $\bar{\beta}(3/2_1^+)=0.13$ and $\bar{\gamma}(3/2_1^+)=40^\circ$. Following the the strategy of Ref.~\cite{Bally22a}, we then fitted the parameters of a deformed Woods-Saxon density profile, to obtain a new state-of-the-art microscopically-motivated input for the simulation of high-energy $^{197}$Au+$^{197}$Au collisions. In terms of radial profile parameters, our result corrects to some extent the widely- and incorrectly-employed charge-density parametrization, which has in particular a too large skin thickness. For future precision phenomenological studies of $^{197}$Au+$^{197}$Au collisions, especially in view of the upcoming sPHENIX program at the BNL RHIC, it will be crucial to implement realistic properties of the point-nucleon density in Monte Carlo simulations. This includes as well implementing an appropriate triaxiality, of order $45^\circ$, for gold ions. Our estimates indicate that this magnitude of the triaxiality does impact the final state in a significant way, and we expect future theoretical work to be able to cleanly isolate such a contribution from the data. As an outlook, we emphasize that measurements of the third centered moment (skewness) of the distribution of $\langle p_t \rangle$ \cite{Giacalone21a} provide additional and independent information about $\gamma^{\rm WS}$ \cite{Jia:2021qyu}, and can be used in conjunction with hydrodynamic simulations to further test our prediction for this parameter.

%
%
\begin{acknowledgements}
We thank Chunjian Zhang for help with the entropy-to-multiplicity conversion used in Fig.~\ref{fig:G1}, and Wouter Ryssens for useful discussions.
This project has received funding from the European Union’s Horizon 2020 research and innovation programme under the Marie Sk\l{}odowska-Curie grant agreement No.~839847. 
M.B.~acknowledges support by the Agence Nationale de la Recherche, France, under grant No.~19-CE31-0015-01 (NEWFUN). G.G.~is funded by the Deutsche Forschungsgemeinschaft (DFG, German Research Foundation) under Germany's Excellence Strategy EXC2181/1-390900948 (the Heidelberg STRUCTURES Excellence Cluster), within the Collaborative Research Center SFB1225 (ISOQUANT, Project-ID 273811115).
The calculations were performed by using HPC resources from CIEMAT (Turgalium), Spain (FI-2021-3-0004, FI-2022-1-0004).
\end{acknowledgements}

%
%
\interlinepenalty=10000
\bibliography{biblio.bib, biblio_ions.bib}
%
%
\end{document}